\documentclass[twocolumn]{aastex63}

\usepackage{amsmath}

\begin{document}

\title{Rarefied Broad-Line Regions in Active Galactic Nuclei: \\ 
Anomalous Responses in Reverberation Mapping and Implications for Weak Emission-Line Quasars}

\author[0000-0002-5830-3544]{Pu Du}
\affiliation{Key Laboratory for Particle Astrophysics, 
Institute of High Energy Physics, Chinese Academy of Sciences, 
19B Yuquan Road, Beijing 100049, People's Republic of China}

\author{Shuo Zhai}
\affiliation{Key Laboratory for Particle Astrophysics, 
Institute of High Energy Physics, Chinese Academy of Sciences, 
19B Yuquan Road, Beijing 100049, People's Republic of China}

\author[0000-0001-9449-9268]{Jian-Min Wang}
\affiliation{Key Laboratory for Particle Astrophysics, 
Institute of High Energy Physics, Chinese Academy of Sciences, 
19B Yuquan Road, Beijing 100049, People's Republic of China}
\affiliation{National Astronomical Observatories of China, 
Chinese Academy of Sciences, 20A Datun Road, 
Beijing 100020, People's Republic of China}
\affiliation{School of Astronomy and Space Science, 
University of Chinese Academy of Sciences, 
19A Yuquan Road, Beijing 100049, People's Republic of China}

\begin{abstract}
    Reverberation mapping (RM) is a widely-used method for probing the physics
    of broad-line regions (BLRs) in active galactic nuclei (AGNs). There are
    increasing preliminary evidences that the RM behaviors of broad emission
    lines are influenced by BLR densities, however, the influences have not been
    investigated systematically from theoretical perspective. In the present
    paper, we adopt locally optimally emitting cloud model and use CLOUDY to
    obtain the one-dimensional transfer functions of the prominent UV and
    optical emission lines for different BLR densities. We find that the
    influences of BLR densities to RM behaviors have mainly three aspects.
    First, rarefied BLRs (with low gas densities) may show anomalous responses
    in RM observations. Their emission-line light curves inversely response the
    variations of continuum light curves, which may have been observed in some
    UV RM campaigns. Second, the different BLR densities in AGNs may result in
    correlations between the time lags and equivalent widths of emission lines,
    and may contribute to the scatters of the radius-luminosity relationships.
    Third, the variations of BLR densities may explain the changes of time lags
    in individual objects in different years. Some weak emission-line quasars
    (WLQs) are probably extreme cases of rarefied BLRs. We predict that their RM
    observations may show the anomalous responses.
\end{abstract}


\section{Introduction} 
\label{sec:intro}

In the UV and optical spectra of active galactic nuclei (AGNs), strong broad
emission lines (BELs) with velocity widths of $10^3$ -- $10^4$ km/s are one of
the most prominent features. BELs originate from the photoionization of the
gaseous clouds in the broad-line regions (BLRs) driven by the continuum
radiation from the accretion disks around the central supermassive black holes
(SMBHs). The fluxes or equivalent widths (EWs) of BELs are determined by, e.g.,
the strength of ionizing radiation, the amount of BLR gas, or the reprocessing
coefficients, while their line profiles are controlled by the bulk motions of
BLR gas governed by the gravitational potential of SMBHs. Understanding the BELs
and the underlying BLR physics is crucial to revealing the origin and evolution
of AGNs. 

Reverberation mapping (RM) is a classic tool for the investigation of BLR
properties in AGNs and the mass measurement of their SMBHs \citep{Bahcall1972,
Blandford1982}. It measures the time delays of emission lines (emission-line
light curves) relative to the variation of the continuum (continuum light curve),
and has been applied to more than a hundred of AGNs in the past three decades
\citep[e.g.,][]{Peterson1998, Kaspi2000, Kaspi2021, Bentz2009, Du2014, Du2018,
Grier2017, Lira2018, Yu2021}. RM can probe the photoionization properties of the
material in BLRs based on the response of emission-line flux with respect to the
varying continuum \citep[e.g.,][]{Goad1993, Gilbert2003, Korista2004,
Cackett2006} and diagnose the BLR geometry and kinematics from the response as
a function of velocity \citep[e.g.,][]{Welsh1991, Bentz2009, Denney2010,
Grier2013, Pancoast2014, Du2016b, U2022, Villafana2022}.

The theoretical calculations of RM signals based on photoionization models
started in 1990s. For example, \cite{Goad1993} introduced photoionization models
into the calculations of the responses for different emission lines for the first
time. \cite{Bottorff1997} presented a more sophisticated kinematic model
incorporating photoionization calculations for the one- and two-dimensional
transfer functions of NGC 5548 (focusing on C {\sc iv} emission line).
\cite{Kaspi1999} performed photoionization calculations using  
a pressure-confined model in order to reproduce the light curves of five
emission lines in the RM observations of NGC 5548. \cite{Korista2000} adopted
the locally optimally emitting clouds model \citep{Baldwin1995} and obtained a
better fitting to the UV emission-line light curves of NGC 5548.
\cite{Negrete2014} used CLOUDY and the flux ratios of UV lines to derive the BLR
radii. \cite{Goad2014} investigated the influences to the variation amplitudes
and time delays of emission lines from light-curve durations, sampling rates,
and the time scales of driving continuum variabilities based on photoionization
simulations. More recently, \cite{Guo2020} adopted photoionization models to
explain the behavior of Mg II emission line in RM. \cite{Zhang2021} compared the
BLR sizes measured from RM and spectroastrometry based on photoionization
models. However, all of those calculations payed attention only to the BLRs of
typical AGNs (with typical emission-line EWs, e.g., NGC~5548). 

The EWs of BELs in AGNs have wide distributions, which roughly span more than
one order of magnitude for the primary emission lines like H$\beta$, Mg {\sc
ii}, and C {\sc iv} \citep[e.g.,][]{Boroson1992, Marziani2003, Shen2011}. It
means that the BLR properties, especially the gas content or ionization state,
may be different for different objects. Furthermore, there are non-negligible
populations of AGNs in which even weaker or more rarefied BLRs may reside. One
population is the so-called ``weak emission-line quasars (WLQs)'' in high
redshifts discovered in the past 20 years \citep[e.g.,][]{Fan1999, Anderson2001,
Collinge2005, Diamond-Stanic2009, Plotkin2010, Wu2011, Meusinger2014,
Andika2020}. They are characterized by the significantly weaker Ly$\alpha$
$\lambda1216$ $+$ N {\sc v}$\lambda1240$ and/or C {\sc iv}$\lambda1549$ emission
lines in the UV spectra (in rest frames) than the main population of quasars
\citep[e.g.,][]{Diamond-Stanic2009}. They have high accretion rates but very
weak BELs. There are also some other AGN populations at low redshifts which have
relatively weak BELs, e.g., Seyfert 1.8/1.9 galaxies
\citep[e.g.,][]{Trippe2010}, ``naked'' AGNs \citep[e.g.,][]{Panessa2009}. From
theoretical perspective, the AGNs with relatively weak BELs (hosting rarefied
BLRs) may be at the early stage of BLR evolution \citep[][]{Hryniewicz2010,
Wang2012}, which are probably important to understanding the AGN physics.
However, the RM behaviors of the AGNs with rarefied BLRs (refer to low BLR gas
densities) have not been investigated from theoretical calculation or
systematically from observation. Recent RM observations of UV emission lines
have revealed some anomalous behaviors \citep{Lira2018}. What's surprising is
that they just appear in the objects with weak emission-line EWs \citep[CT320,
CT803, and J224743 in][see its Section 3.3]{Lira2018}. These
UV lines show inverse correlations (negative responses) with the continuum
variations (the emission-line flux goes down/up when the continuum flux
increases/decreases, see more details in Section \ref{sec:ImplicationForRM}).
This phenomenon is different from the ``BLR holiday'' anomaly in the emission
lines of NGC~5548 \citep{Goad2016, Pei2017} and Mrk~817 \citep{Kara2021} found
by the first and second AGN Space Telescope \& Optical Reverberation Mapping
(STORM) programs. During the ``BLR holiday'' period in AGN STORM, the emission
lines decoupled from the continuum variations and showed weak (even no)
correlations, which is different from the inverse correlations found in
\cite{Lira2018} and can be explained by the obscuration from the disk wind
\citep[e.g.,][]{Dehghanian2019}. These thoughts and current situations motivate
us to investigate the RM behaviors of the emission lines in the AGNs with
rarefied BLRs from photoionization calculations. 

Furthermore, the scatter of the famous radius-luminosity ($R$-$L$) relationship
discovered by RM \citep[e.g.,][]{Kaspi2000, Bentz2013, Du2019} is also far from
fully understood. Recent RM campaigns found that the scatter became larger with
more objects (with different properties) been observed. For example,
super-Eddington accreting massive black holes project found that the H$\beta$
lags of the AGNs with high accretion rates are shorter than the prediction of
the $R$-$L$ relationship by factors of $\sim$3-8 \citep{Du2015, Du2016, Du2018}.
Mg {\sc ii} \citep{Martinez-Aldama2020} and C {\sc iv} \citep{DallaBonta2020}
emission lines also show preliminary signs of the similar behaviors. The Sloan
Digital Sky Survey RM project discovered that the H$\beta$ lines of some quasars
with moderate Eddington ratios also show shortened lags \citep{Grier2017}. The
possible explanations for these shortened lags include (1) the self-shadowing
effects of slim accretion disks in super-Eddington AGNs \citep{Wang2014} and (2)
the variation of the spectral energy distribution caused by the spin of BHs
\citep{Wang2014b}. However, it's not yet known whether they are the only drivers
for the scatter of the $R$-$L$ relationship. The influence of the BLR densities
to the time lags has not been investigated systematically for different emission
lines. This is also a main goal of this paper. 

More recently, the RM observations of some objects showed a surprising
phenomenon that their time lags changed significantly in different years,
however the corresponding continuum luminosities were quite similar, e.g.,
NGC~3227 in \cite{Denney2010}, \cite{DeRosa2018}, and \cite{Brotherton2020},
Mrk~817 in \cite{Peterson1998}, \cite{Denney2010}, and \cite{Lu2021}, Mrk~79 in
\cite{Lu2019} and \cite{Brotherton2020}, and PG~0947+396 in different years in
\cite{Bao2022}. Considering that the time lags are mainly determined by the
ionizing continuum and the properties of BLR gas, these observations indicate
that their BLR might probably change with time because their continuum fluxes
kept almost the same. The density of gas is one of the key BLR properties, and
is therefore worthy of investigation.

This paper is organized as follows. The photoionization calculation is described
in Section \ref{sec:photoionization}. A comparison between the EWs obtained from
the photoionization models and the observations from large quasar samples or RM
samples are provided in Section \ref{sec:EW} for different emission lines.
Section \ref{sec:transfer_function} presents the transfer functions for
different BLR density distributions, as well as the correlations between EWs and
time lags. In this section, we demonstrate that the AGNs with rarefied BLRs may
show anomalous RM behaviors. Some discussions are provided in Section
\ref{sec:discussion} (especially the implications for WLQs). Finally, we briefly
summarize in Section \ref{sec:summary}.

\begin{figure*}[!ht]
    \centering
    \includegraphics[width=\textwidth]{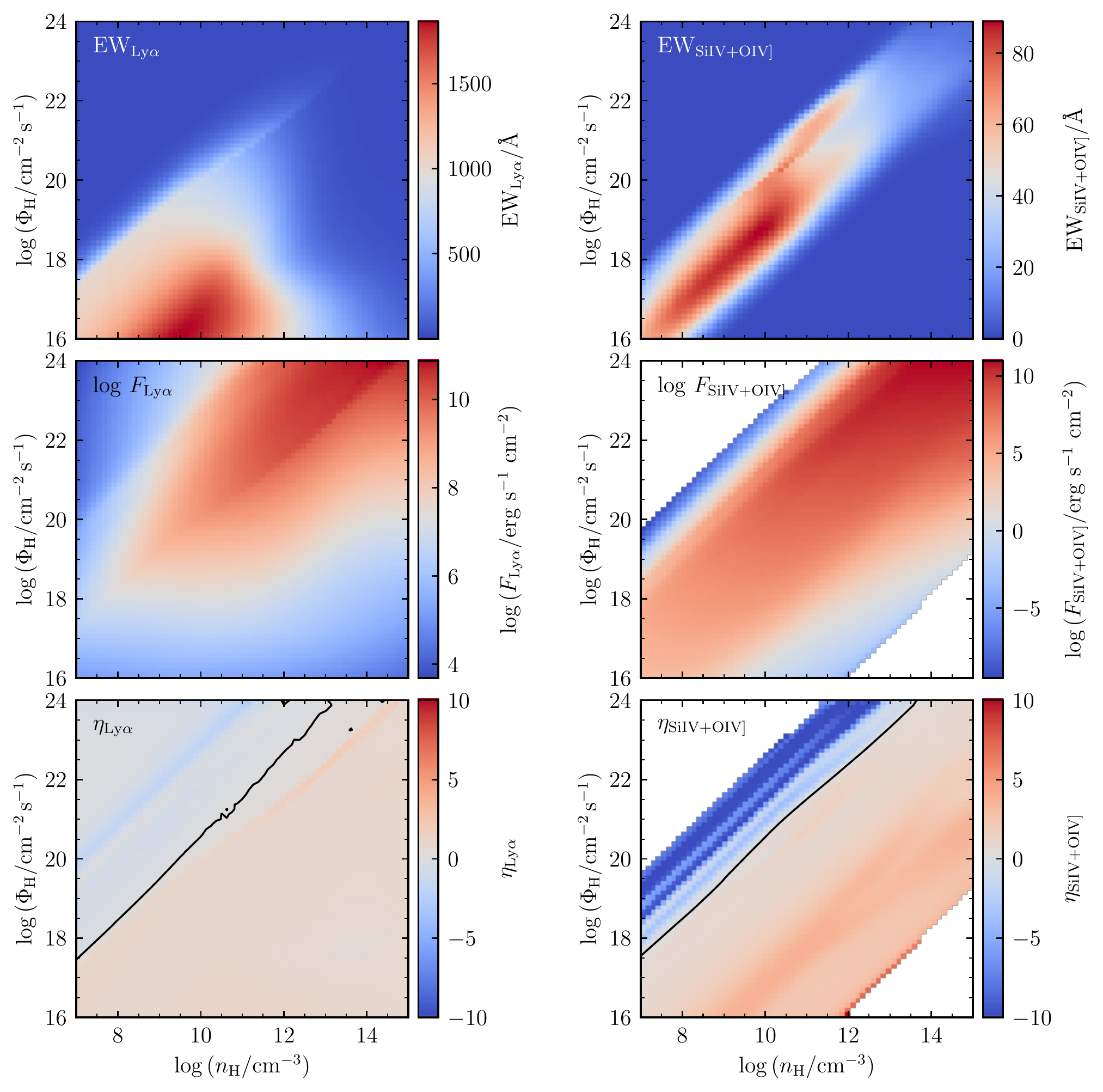}
    \caption{EWs, fluxes, and responsivity of the emission lines for different
    $n_{\rm H}$ and $\Phi_{\rm H}$. It is obvious that the responsivity tends to
    be negative if gas density is lower and ionizing continuum is stronger. The
    solid lines in the lower two panels mark the dividing lines of the positive
    and negative responsivity. \label{fig:ew_flux_eta}}
\end{figure*}

\begin{figure*}
    \figurenum{\ref{fig:ew_flux_eta}}
    \centering
    \includegraphics[width=\textwidth]{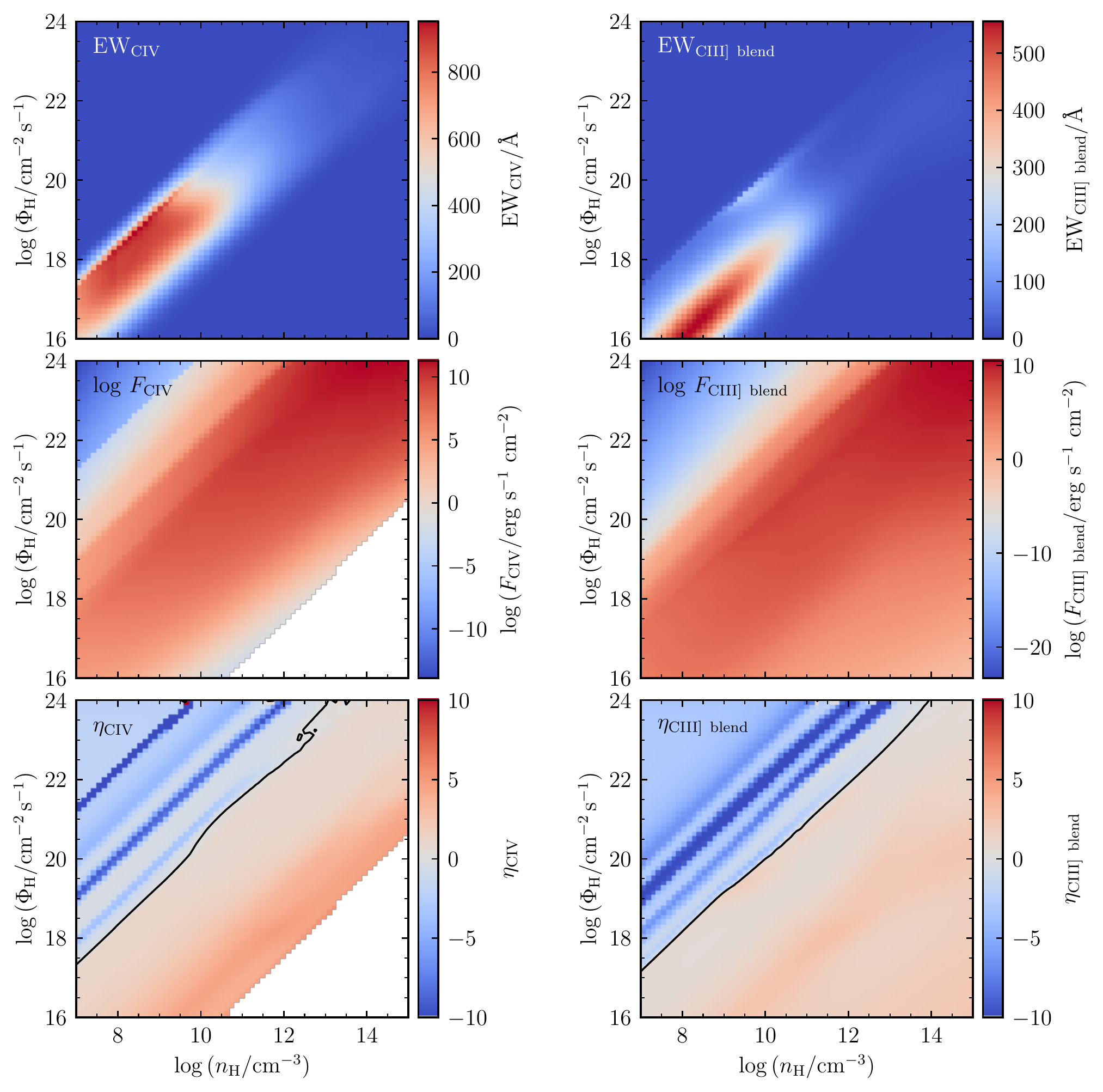}
    \caption{(Continued.)}
\end{figure*}

\begin{figure*}
    \figurenum{\ref{fig:ew_flux_eta}}
    \centering
    \includegraphics[width=\textwidth]{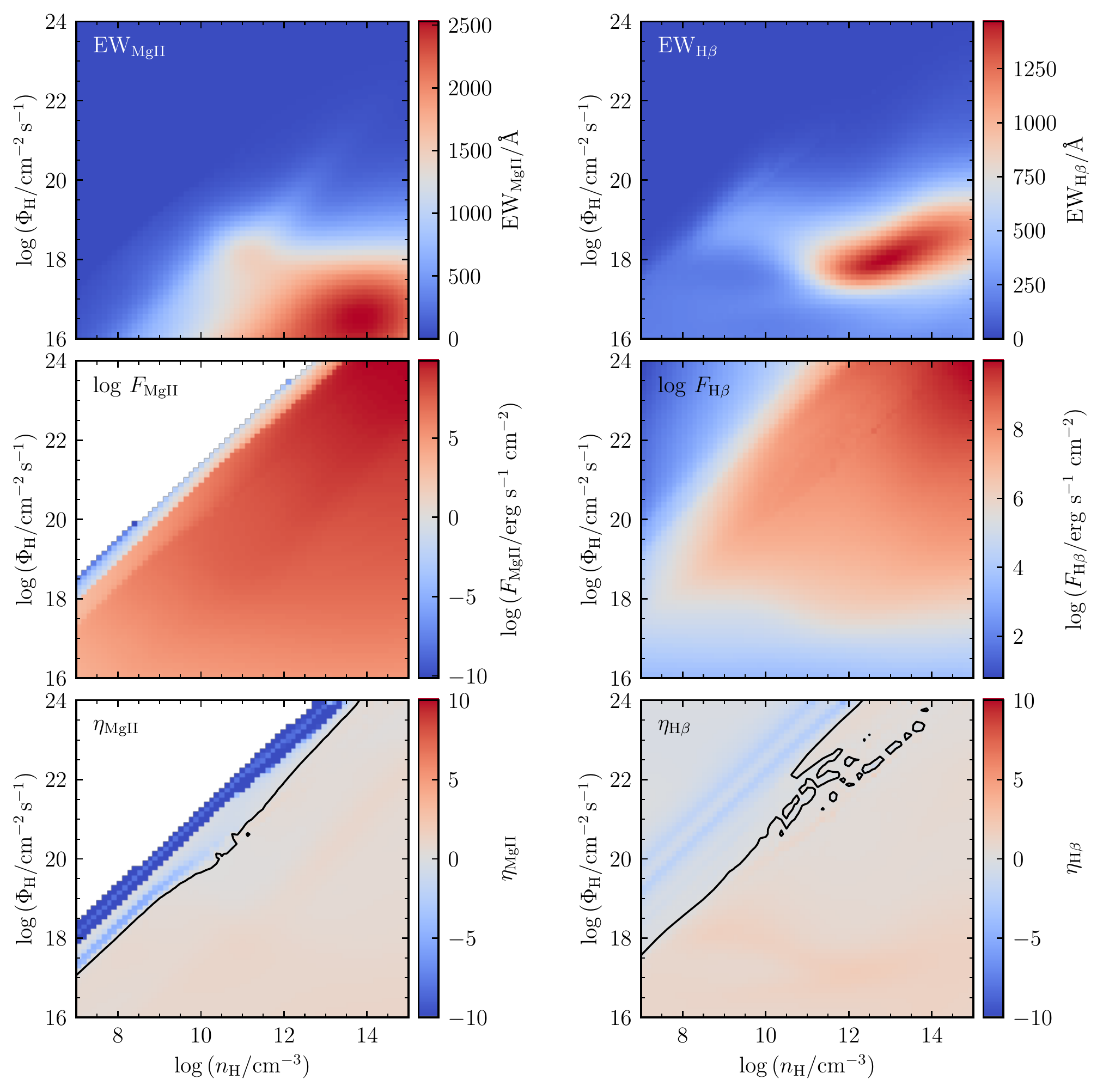}
    \caption{(Continued.)}
\end{figure*}

\section{photoionization Calculation}
\label{sec:photoionization}

Following the pioneering works of photoionization calculations for the
emission-line responses in RM \citep[e.g.,][]{Korista2000, Korista2004}, we
adopt the locally optimally emitting cloud \citep[LOC,][]{Baldwin1995} model and
perform the calculation using CLOUDY v17.02 \citep{Ferland2017}. We generate a
grid of models with gas number density of $7 \leqslant \log\ (n_{\rm H}/{\rm
cm^{-3}}) \leqslant 15$ and surface flux of ionizing photons spanning $16
\leqslant \log\ (\Phi_{\rm H}/{\rm cm^{-2}\, s^{-1}}) \leqslant 24$. The steps
in $n_{\rm H}$ and $\Phi_{\rm H}$ are both 0.125 dex. We assume a simple slab
geometry with column density of $10^{23}\ {\rm cm^{-2}}$ (a standard value, see
\citealt{Netzer2010} and references therein) for the line-emitting entities. The
typical spectral energy distribution (SED) of \cite{Mathews1987} is employed as
the ionizing continuum (more discussion is given in Section
\ref{sec:other_factors}). The metallicity is assumed to be solar abundance.

\begin{figure*}[!ht]
    \centering
    \includegraphics[width=\textwidth]{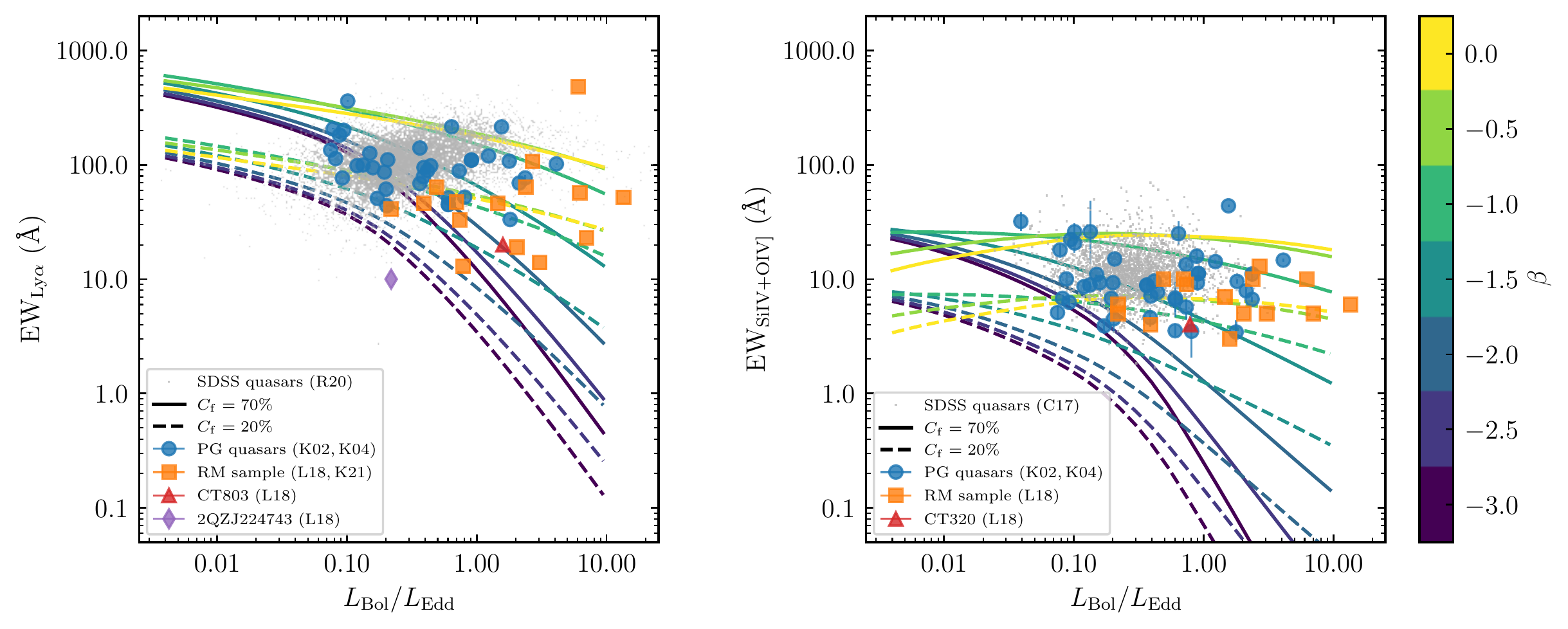}\\
    \includegraphics[width=\textwidth]{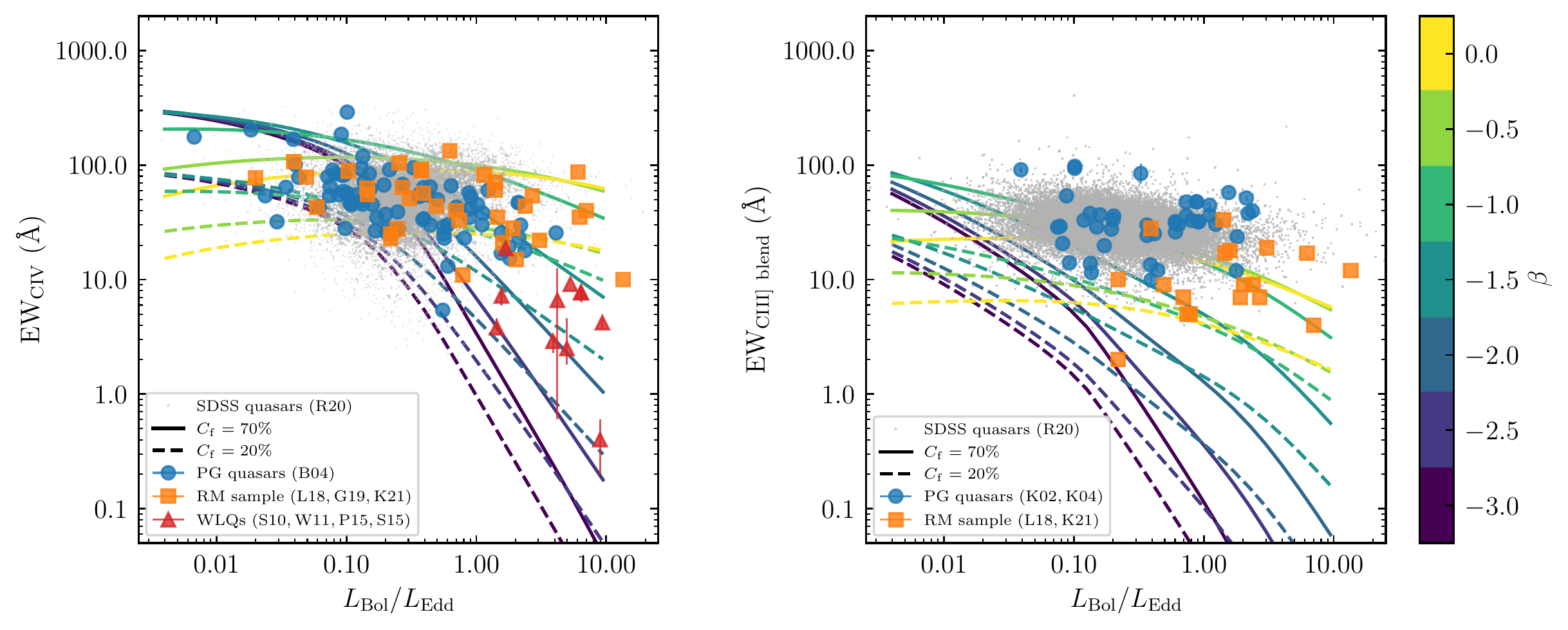}\\
    \includegraphics[width=\textwidth]{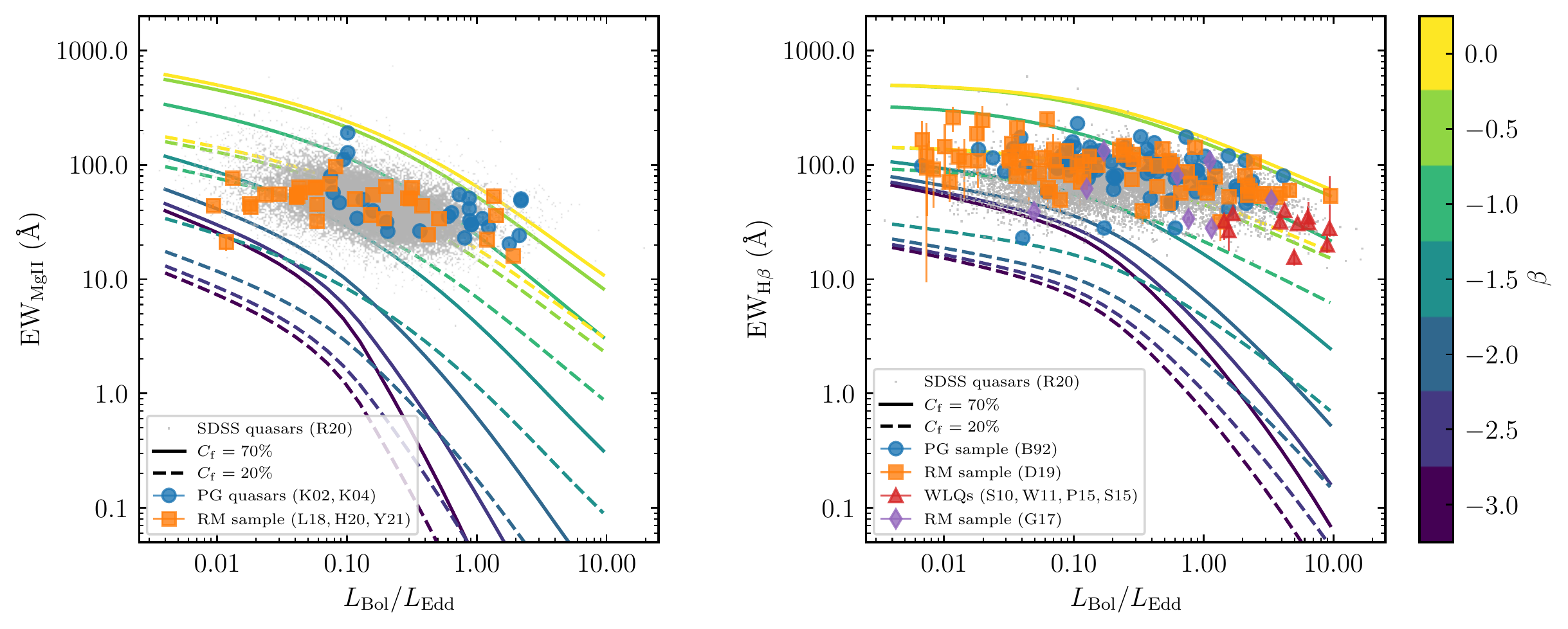}
    \caption{EW vs. Eddington ratio. The lines with different colors are the
    photoionization models with different $\beta$. The solid and dashed lines
    are the results for $C_{\rm f}=70\%$ and $20\%$, respectively. The
    observational points with different symbols in different colors are the
    samples of RM AGNs, PG quasars, SDSS quasars, and WLQs, which are overlapped
    for comparison. \label{fig:ew_beta}}
\end{figure*}

We focus on the emission lines that frequently presented in the studies of UV
\citep[e.g.,][]{Clavel1991, Grier2019, Kaspi2021, Lira2018, Yu2021} and optical
RM observations \citep[e.g.,][]{Peterson1998, Kaspi2000, Du2014, Grier2017,
U2022}. They are Ly$\alpha$ $\lambda1216$, Si {\sc iv}$+$O {\sc iv}]
$\lambda1400$ blend, C {\sc iv} $\lambda1549$ doublet, C {\sc iii}]
$\lambda1909$ blend, Mg {\sc ii} $\lambda2798$ doublet, and H$\beta$
$\lambda4861$ emission lines. There are many other emission lines (Al {\sc iii}
$\lambda\lambda1855,1863$, Si {\sc iii} $\lambda\lambda1883,1892$, and Fe {\sc
iii} $\lambda\lambda1895, 1914, 1926$) that are seriously blended with C {\sc
iii}] \citep[see, e.g.,][]{Negrete2012, Temple2020}. They are also added into
the flux of C {\sc iii}] blend. 

In order to demonstrate the responses of the BLR clouds to variation of the
continuum for different $\Phi_{\rm H}$ and $n_{\rm H}$, we adopt the definition
of responsivity ($\eta$) in \cite{Korista2004}. It is defined as
\begin{equation}\label{eqn:eta}
    \eta = \frac{d \log F_{\ell}}{d \log \Phi_{\rm H}} 
         = \frac{d \log {\rm EW}}{d \log \Phi_{\rm H}} + 1.
\end{equation}
The responsivities of the BLR clouds with different $n_{\rm H}$ and $\Phi_{\rm
H}$ obtained from the photoionization grid, as well as the corresponding EWs and
fluxes, are shown in Figure \ref{fig:ew_flux_eta}. In each panel of the
responsivity in Figure \ref{fig:ew_flux_eta}, a solid line is added as the
division of positive and negative values. It is obvious that the responsivity is
negative if the gas density is low and the surface flux of the ionizing photons
is high (in the upper-left corners of the corresponding panels in Figure
\ref{fig:ew_flux_eta}). This motivates us to consider that the emission-line
responses may gradually become negative if the weights of low-density clouds in
BLRs increase.

Integrating the line emission over the grid, we can obtain the luminosities of
the emission lines. Following \cite{Baldwin1995} and \cite{Bottorff2002}, the
emission-line luminosity can be obtained by
\begin{equation} \label{eq:Lline}
    L_{\ell} \propto \int_{n_{\rm min}}^{n_{\rm max}}\!\!\int_{r_{\rm in}}^{r_{\rm out}} r^2 F_{\ell}(n, r) f(r) g(n) dn dr,
\end{equation}
where $F_{\ell}(n, r)$ is the line flux of the clouds with gas density $n$ at
radius $r$, $f(r)$ is the covering fraction, and $g(n)$ is the gas density
distribution. If the ionizing luminosity $L_{\rm ion}$ is given, the radius and the flux of
ionizing photons are connected through $\Phi_{\rm H} \propto L_{\rm ion} / 4 \pi r^2$. We
assume that both of $f(r)$ and $g(n)$ are simply power laws, namely $f(r)
\propto r^{\Gamma}$ and $g(n) \propto n^{\beta}$, where $\Gamma$ and $\beta$ are
two indexes. The goal of the present paper is to investigate the potential
observational characteristics in RM if the BLR densities change. We fix $\Gamma$
as a constant and keep $\beta$ as a free parameter. As an AGN prototype, the
optimal $\Gamma$ in NGC~5548 was suggested to be in the range of $-1.4 < \Gamma
< -1.0$ \citep{Korista2004}. For simplicity, we adopt $\Gamma = -1.0$ in this
work. 

For typical AGNs, $\beta = -1.0$ is an acceptable assumption \citep{Baldwin1995,
Baldwin1997, Korista2004, Guo2020}. Smaller $\beta$ means that most of the BLR
clouds have lower gas densities (rarefied BLRs), and larger $\beta$ stands for
the BLRs with higher densities (dense BLRs). Here we set $\beta$ from $-3.0$ to
$0.0$ to check the influences of gas density to the response behaviors of
different emission lines in RM observations. For UV emission lines, we adopt
$n_{\rm min} = 10^8\, {\rm cm^{-3}}$ and $n_{\rm max} = 10^{12}\, {\rm cm^{-3}}$
because lower density may lead to some unobserved forbidden lines and higher
density makes, e.g., C {\sc iv} thermalized \citep{Korista2000, Korista2004}.
However, hydrogen recombination line can emit efficiently even in much higher
density (see Figure \ref{fig:ew_flux_eta} and \citealt{Korista1997}). We loose
$n_{\rm max}$ to $10^{15}\, {\rm cm^{-3}}$ for the calculation of the H$\beta$
emission line. 

To obtain the emission-line luminosity (Eqn \ref{eq:Lline}), the inner and outer
radii ($r_{\rm in}$ and $r_{\rm out}$) of BLRs are required. From Figure
\ref{fig:ew_flux_eta}, the most efficiently emitting clouds are located in
relatively a narrow range of $\Phi_{\rm H}$. In the present paper, we pay more
attention to the influences of different Eddington ratios than those of the BH
masses. For a given BH mass $M_{\bullet}$, we adopt $r_{\rm in} =
\sqrt{Q_{0.1}/4\pi\Phi_{\rm H}^{\rm max}}$ and $r_{\rm out} =
\sqrt{Q_{0.1}/4\pi\Phi_{\rm H}^{\rm min}}$, where $\Phi_{\rm H}^{\rm min} =
10^{18}$ cm$^{-2}$~s$^{-1}$ and $\Phi_{\rm H}^{\rm max} = 10^{22}$
cm$^{-2}$~s$^{-1}$. $Q_{0.1}$ is the total number of ionizing photons if the
Eddington ratio $L_{\rm Bol}/L_{\rm Edd}=0.1$, where $L_{\rm Bol}=\kappa
L_{5100}$ is the bolometric luminosity, $\kappa$ is the bolometric correction
factor, and $L_{\rm Edd}=1.26\times10^{38} M_{\bullet} \, {\rm erg\, s^{-1}}$ is
the Eddington luminosity. We simply adopt $\kappa=10$ here
\citep[e.g.,][]{Richards2006}, but caution that $\kappa$ may depend on the
accretion rate or SMBH mass \citep[e.g.,][]{Jin2012}. The $\Phi_{\rm H}^{\rm
min}$ and $\Phi_{\rm H}^{\rm max}$ values we adopted here are just corresponding
to $r_{\rm in} = 0.1 r_{\rm BLR}^{\rm typical}$ and $r_{\rm out} = 10 r_{\rm
BLR}^{\rm typical}$, where $r_{\rm BLR}^{\rm typical}$ is the typical BLR size
of the AGNs with $L_{\rm Bol}/L_{\rm Edd}=0.1$ calculated from the classical
radius-luminosity (R-L) relationship of $\log (r_{\rm BLR}/{\rm lt\!-\!days}) =
1.53 + 0.53 \log \ell_{44}$ in \cite{Bentz2013}, where
$\ell_{44}=L_{5100}/10^{44}\ {\rm erg\ s^{-1}}$ is the monochromatic luminosity
at 5100\AA. $L_{\rm Bol}/L_{\rm Edd}=0.1$ is roughly the central value of
Eddington ratios for nearby Seyfert galaxies and high-redshift quasars ($0.01
\lesssim L_{\rm Bol}/L_{\rm Edd} \lesssim 1$, e.g., \citealt{Boroson1992,
Marziani2003, Shen2011, Du2019, Wu2015}, and see Figure \ref{fig:ew_beta}), so
the corresponding $r_{\rm BLR}^{\rm typical}$ can be regarded as a typical
value. It should be noted that $L_{\rm Bol}/L_{\rm Edd}=0.1$
here is only used for determining a relatively reasonable $r_{\rm in}$ and
$r_{\rm out}$. For higher (or lower) ionizing luminosity $L_{\rm ion}$ (or
Eddington ratio $L_{\rm Bol}/L_{\rm Edd}$), the most efficiently emitting radius
of the line region will spontaneously increases (or decreases) in accordance
with $r \propto L_{\rm ion}^{1/2}$. This is known as the physical interpretation
for the R-L relationship established from the RM campaigns
\citep[e.g.,][]{Kaspi2000, Bentz2013}. Such boundary assumption makes the EW
calculation scale-free from BH mass. The dynamic range of radius adopted here is
generally large enough for the span of Eddington ratio. We have checked that
slightly larger or smaller $r_{\rm in}$ and $r_{\rm out}$ do not influence the
general results of the present paper.

\section{Equivalent Widths}
\label{sec:EW}

In Figure \ref{fig:ew_beta}, we show the dependences of EW on $L_{\rm
Bol}/L_{\rm Edd}$ for the emission lines obtained from our photoionization
model. The cases with overall covering factors of $C_{\rm f}=20\%$ and $70\%$
are demonstrated. In general, smaller $\beta$ (more rarefied BLR) tends to show
weaker EWs. But at the low Eddington-ratio ends of Si {\sc iv}$+$O
{\sc iv}], C {\sc iv}, and C {\sc iii}], large $\beta$ (dense BLR) also produces
small EWs. 

In order to compare our theoretical calculations with observations
and to generally determine the ranges of $\beta$ for different emission lines,
we collect the emission-line EWs and $L_{\rm Bol}/L_{\rm Edd}$ from the following
samples: (1) the objects from recent RM campaigns or compilations
\citep{Grier2017, Lira2018, Du2019, Homayouni2020, Yu2021, Kaspi2021}, (2) the
H$\beta$ \citep{Boroson1992} and UV \citep{Kuraszkiewicz2002, Kuraszkiewicz2004}
emission lines of PG quasars, (3) the emission lines of the quasar samples
\citep{Calderone2017, Rakshit2020} from Sloan Digital Sky Survey (SDSS), and (4)
the WLQs from \cite{Shemmer2010}, \cite{Wu2011}, and \cite{Plotkin2015}. 

For the RM objects, we can easily calculate their Eddington ratios based on the BH masses
obtained from the time lags \citep{Grier2017, Lira2018, Du2019, Homayouni2020,
Yu2021, Kaspi2021}. All of the objects in PG sample have H$\beta$ observations. 
For the PG quasars without RM measurements, the single-epoch
BH masses estimated from their H$\beta$ lines can be used. We employ the simple
virial relation to determine their BH masses, namely 
\begin{equation}
M_{\bullet} = \frac{f V_{\rm H\beta}^2 r_{\rm H\beta}}{G}, 
\end{equation}
where $V_{\rm H\beta}$ is the FWHM of H$\beta$ line, $G$ is the gravitational
constant, and $f$ is the virial factor. We simply adopt $f=1$ in this paper
\citep[e.g.,][]{Woo2015}. As aforementioned, the recent works
\citep[e.g.,][]{Du2015, Du2018, Du2019} discovered that BLR radius (measured
from H$\beta$) depends on accretion rate and suggested a new scaling relation
for $r_{\rm H\beta}$, which includes the relative strength of Fe {\sc ii} lines
as a new parameter. We calculate the BH masses of the PG objects, which have
both H$\beta$ and Fe {\sc ii} measurements, using the new scaling relation of 
\begin{equation}
\log (r_{\rm H\beta}/{\rm lt\!-\!days}) = 1.65 + 0.45 \log \ell_{44} - 0.35 R_{\rm Fe}
\end{equation} 
in \cite{Du2019}, where $R_{\rm Fe}$ is the flux ratio between Fe {\sc ii} and
H$\beta$ lines. For the SDSS quasars with low redshifts which have H$\beta$ observations, the
single-epoch BH masses based on the new scaling relation are also used. For the
SDSS quasars with high redshifts (with only UV lines), their BH masses and
Eddington ratios are based on the classic single-epoch BH mass estimators
\citep[see][]{Calderone2017, Rakshit2020}.  It should be noted that the
Eddington ratios of the SDSS quasars with high redshifts (for the UV lines) may
be underestimated to some extent because their BH masses are obtained based on
the classic single-epoch BH mass estimators and the shortening effects of the
time lags \citep[e.g.,][]{Martinez-Aldama2020, DallaBonta2020} have not been
taken into account. For the WLQs, we only select the objects that have both of C {\sc iv} and
H$\beta$ measurements \citep{Shemmer2010, Wu2011, Plotkin2015}. An advantage of
selecting these WLQs is that, we can obtain relatively good $M_{\bullet}$
estimates for these objects using the new scaling relation \citep{Du2019},
considering that the H$\beta$ emission lines in WLQs are not significantly
different from those of normal AGNs (see more details in Section
\ref{sec:WLQ_Hb_CIV}).

In Figure \ref{fig:ew_beta}, the emission lines show the Baldwin effects
\citep[e.g.,][]{Baldwin1977} but with different slopes and significances. In
general, our photoionization models can cover the distributions of the
observational points, which validates the settings of the parameters used in our
calculations. Comparing the EW distributions of the current RM samples with
those of the other samples, it is obvious that there is still room to improve
the completeness of the RM samples. For example, the current RM sample of C {\sc iv}
lacks low-EW objects, especially WLQs, and the H$\beta$ RM sample 
lines also shows a little bias toward high EWs. On the contrary, more RM
observations of Ly$\alpha$, Si {\sc iv}$+$O {\sc iv}], and C {\sc III}] emission
lines are needed for high-EW AGNs. Moreover, it is obvious that the objects,
which showed anomalous RM behaviors in \cite{Lira2018} (CT320, CT803, and
2QZJ224743), are located at the lower EW ends of the corresponding panels in
Figure \ref{fig:ew_beta}. It implies that their anomalous behaviors may probably
connect with weak EWs. 

Figure \ref{fig:ew_beta} can give a general constraints to the parameter $\beta$
in the context of the photoionization calculations in the present paper. For
example, we can determine that the $\beta$ parameter of WLQs roughly ranges from
$-2.0$ to $-1.0$ for C {\sc iv} and from $-0.7$ to $0.0$ for H$\beta$ for the
case with $C_{\rm f}=20\%$, and from $-2.5$ to $-1.5$ for C {\sc iv} and from
$-1.2$ to $-0.5$ for H$\beta$ for the case with $C_{\rm f}=70\%$, respectively.
It is noted that the observational distributions of the C {\sc III}] blend in
Figure \ref{fig:ew_beta} depart slightly from the photoionization calculations.
One probable reason is the relatively low abundance (solar abundance) we adopted
here (e.g., \citealt{Snedden1999}, also see \citealt{Panda2018, Panda2019,
Sniegowska2021}). We mainly focus on the influence of the BLR densities to the
RM behaviors. The influence from abundance will be discussed in a future
separate paper (see also Section \ref{sec:other_factors} and
Appendix \ref{sec:app}).

\begin{figure*}
    \centering
    \includegraphics[width=\textwidth]{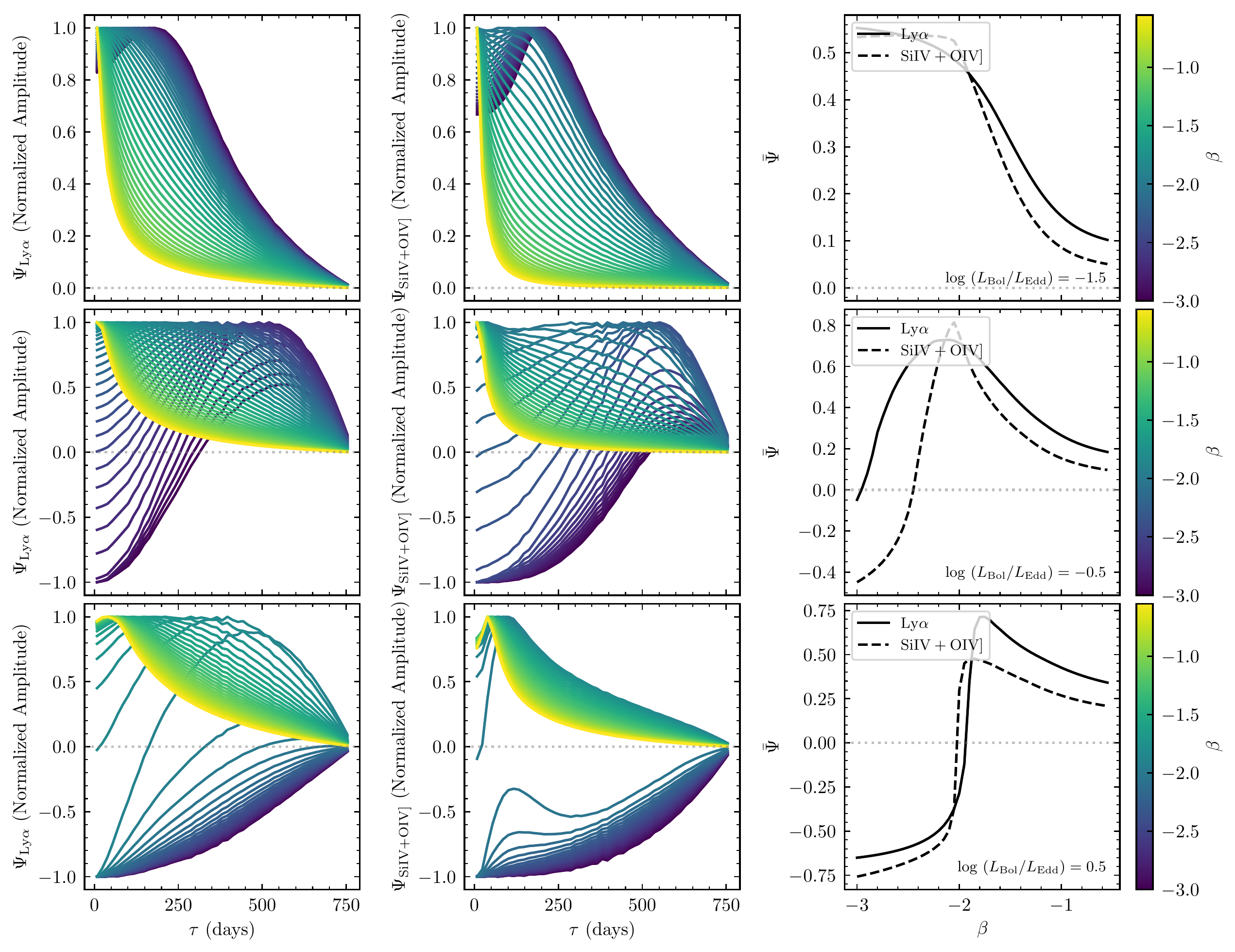}
    \caption{One-dimensional transfer functions $\Psi$ of the emission lines for
    different $\beta$. The transfer functions are color coded by the $\beta$
    parameter. $\bar{\Psi}$ is the average of the transfer function. The cases
    with $\log (L_{\rm Bol}/L_{\rm Edd})=-1.5$, $-0.5$ and $0.5$ are shown,
    respectively. The emission lines tend to show negative responses to the
    varying continuum in RM observations if $\beta$ is smaller.
    \label{fig:transfer_functions}}
\end{figure*}

\begin{figure*}
    \figurenum{\ref{fig:transfer_functions}}
    \centering
    \includegraphics[width=\textwidth]{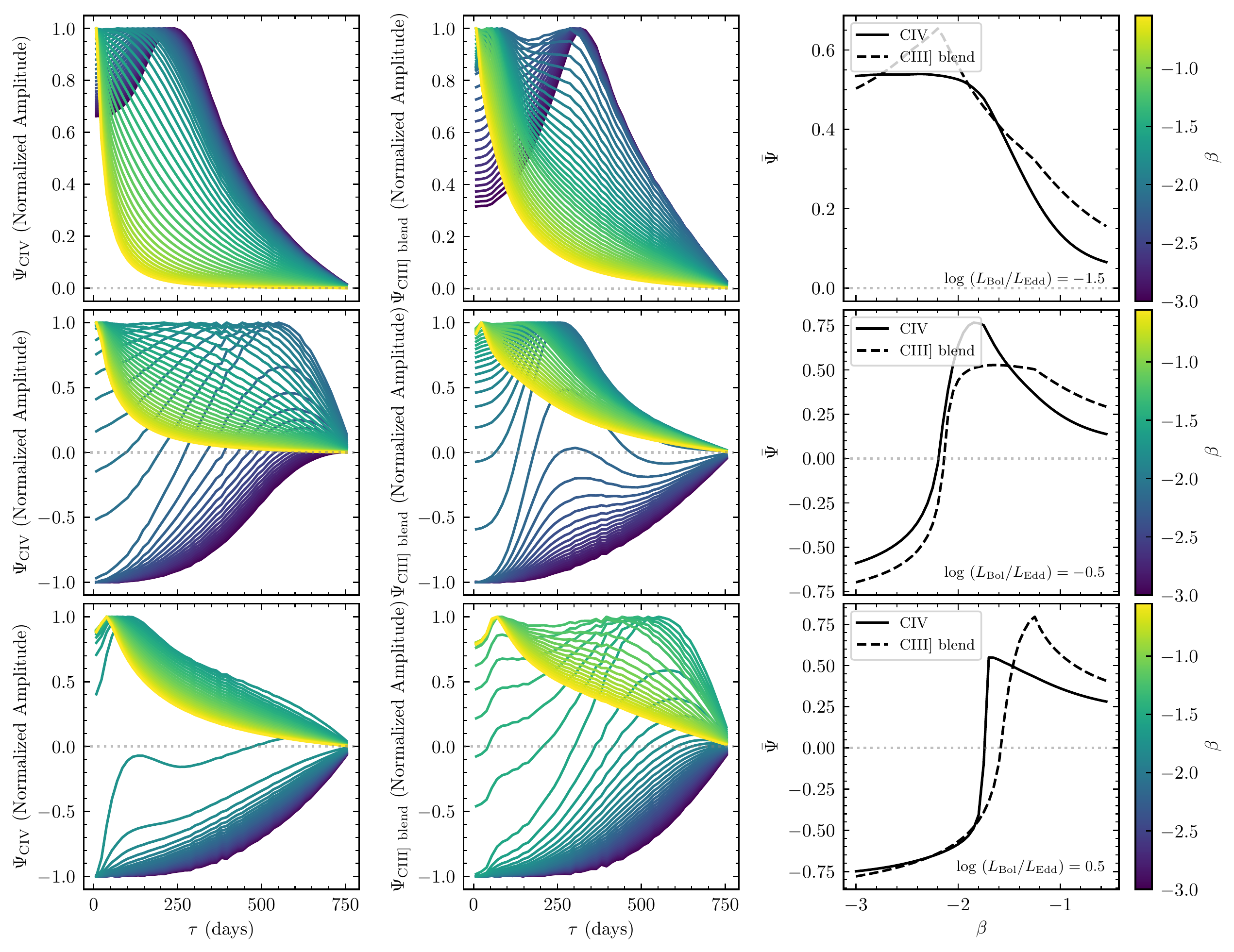}
    \caption{(Continued.)}
\end{figure*}

\begin{figure*}
    \figurenum{\ref{fig:transfer_functions}}
    \centering
    \includegraphics[width=\textwidth]{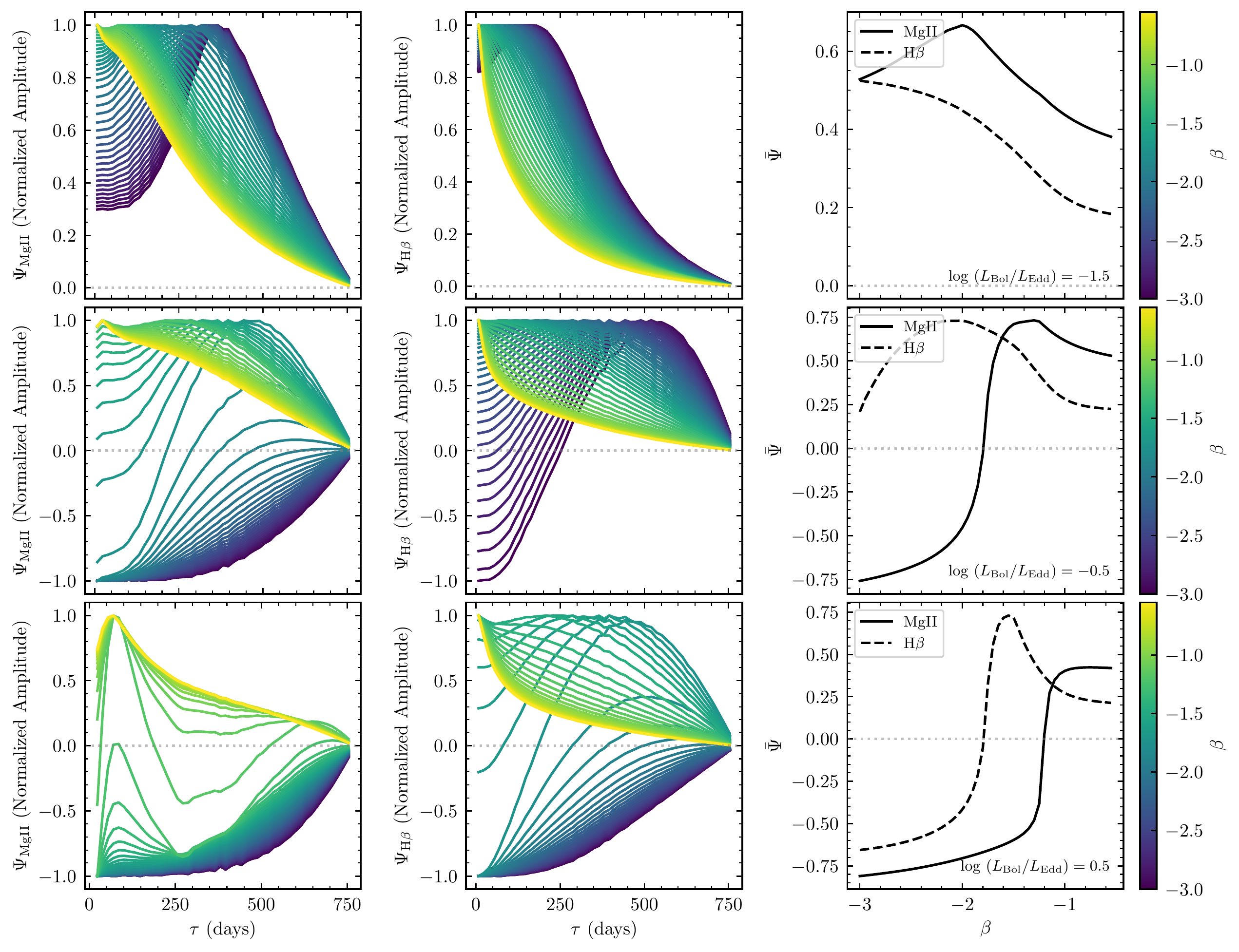}
    \caption{(Continued.)}
\end{figure*}

\section{Transfer function}
\label{sec:transfer_function}

In RM, the delayed response of an emission line to the varying continuum can be
characterized by the transfer function \citep{Blandford1982} in the form of
\begin{equation} \label{eqn:rm}
    \Delta F_{\ell}(t) = \int_{-\infty}^{+\infty} \Delta F_{\rm c}(t^{\prime}) \Psi(t - t^{\prime}) dt^{\prime},
\end{equation}
where $\Delta F_{\rm c}(t)$ and $\Delta F_{\ell}(t)$ are the variations of the
continuum and emission-line fluxes, and $\Psi(t)$ is the one-dimensional
transfer function. Transfer function $\Psi$ connects the emission-line light
curve with the continuum variation \citep{Blandford1982}, and can be easily
obtained from RM observations by different algorithms and softwares, e.g., the
maximum entropy method \citep[e.g.,][]{Krolik1991, Horne1991}, JAVELIN
\citep{Zu2011}, MICA \citep{Li2016}, and Pixon \citep{Li2021}. Some examples of
the one-dimensional transfer functions reconstructed from RM observations can be
found in, e.g., \cite{Grier2013}, \cite{Williams2018}, and \cite{Bao2022}. From
the photoionization grid in Section \ref{sec:photoionization}, we can calculate
the one-dimensional transfer functions by 
%
\begin{multline}
    \Psi(\tau) \propto \int_{r_{\rm in}}^{r_{\rm out}} \int_{0}^{\pi} \int_{0}^{2\pi}
                \eta(r) f(r) F_{\ell}(r) r^2 \sin \theta \\
                 \times \delta\left[\tau - \frac{r + \vec{r} \cdot \vec{n}_{\rm obs}}{c}\right] dr d\theta d\phi,
\end{multline}
%
where
\begin{equation}
    F_{\ell}(r) = \int_{n_{\rm min}}^{n_{\rm max}} F_{\ell}(n, r) g(n) dn,
\end{equation}
is the emission-line flux at radius $r$ obtained by integrating $F(n, r)$ over
density $n$, 
\begin{equation}
    \eta(r) = -0.5 \frac{d \log F_{\ell}(r)}{d \log r}
\end{equation}
is the responsivity function at radius $r$ derived from Eqn (\ref{eqn:eta}),
$\tau$ is the time, $\vec{n}_{\rm obs}$ is the line of sight, $c$ is the speed
of light, and ($\theta$, $\phi$) are the angles of the spherical coordinates.
The purpose of this paper is to investigate the influence of BLR densities to
the transfer functions. Therefore, for simplicity, we assume that the BLR
geometry is spherically symmetric.

\begin{figure*}
    \centering
    \includegraphics[width=0.48\textwidth]{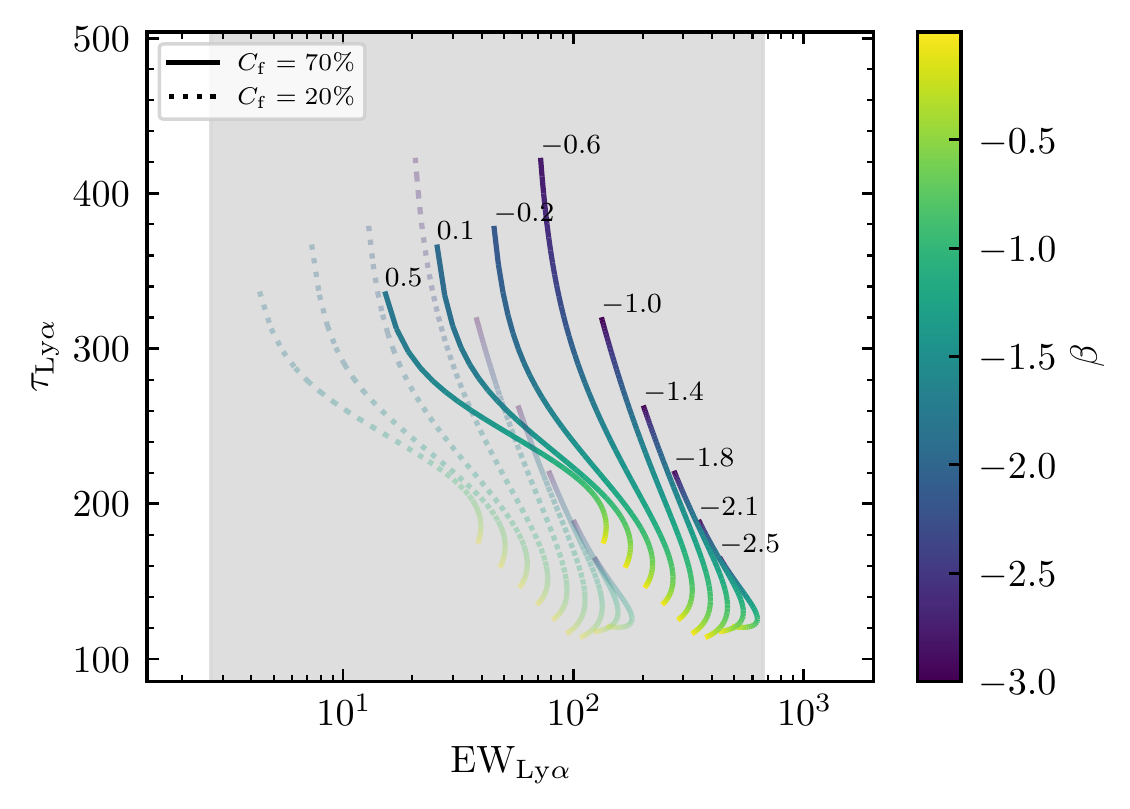}
    \includegraphics[width=0.48\textwidth]{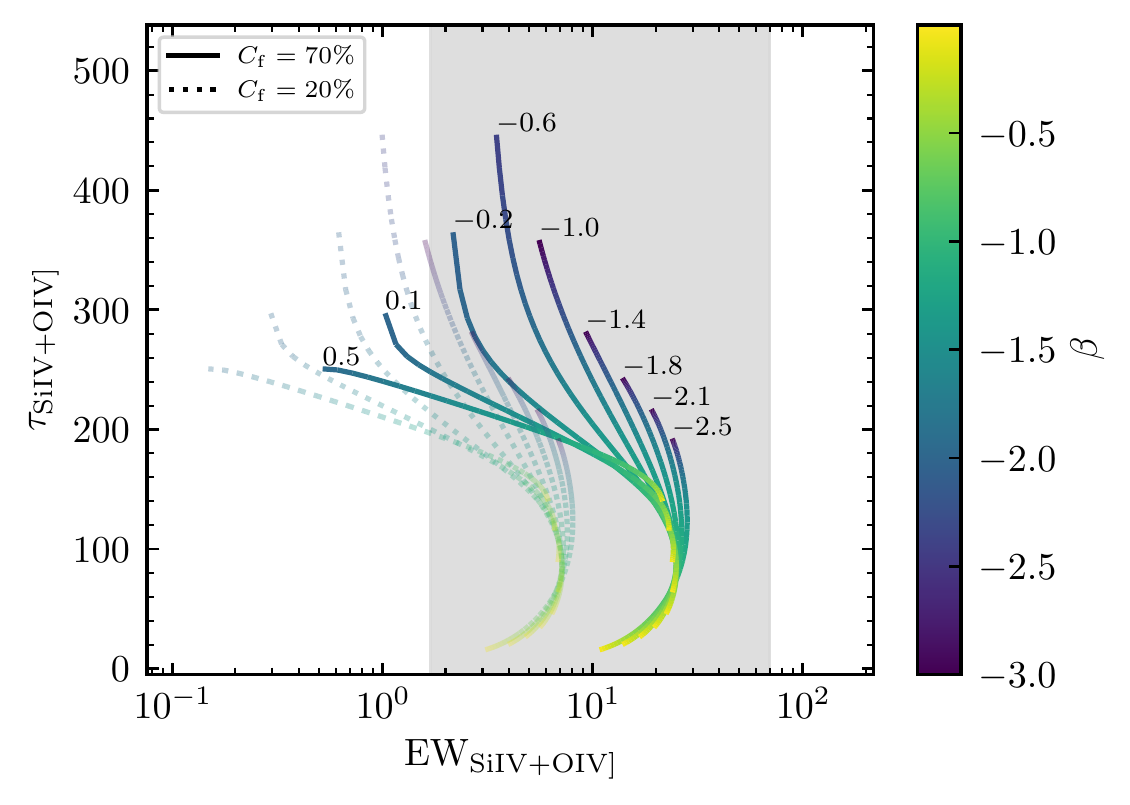} \\
    \includegraphics[width=0.48\textwidth]{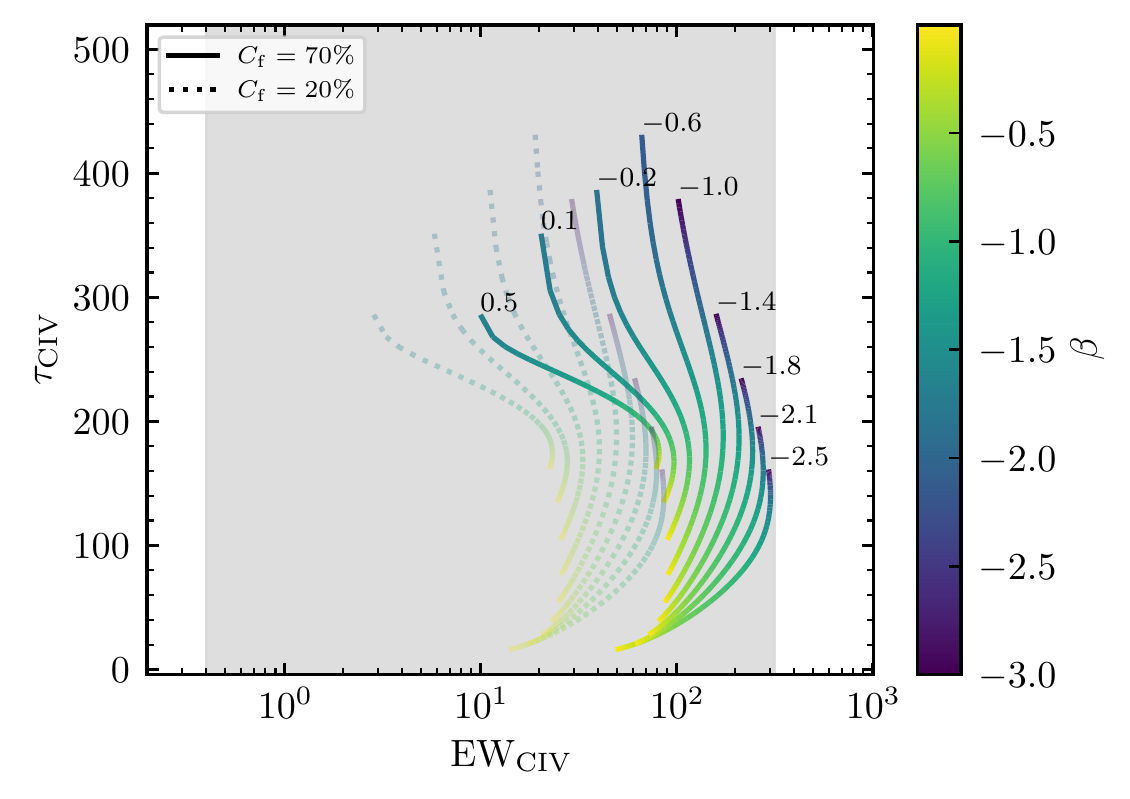}
    \includegraphics[width=0.48\textwidth]{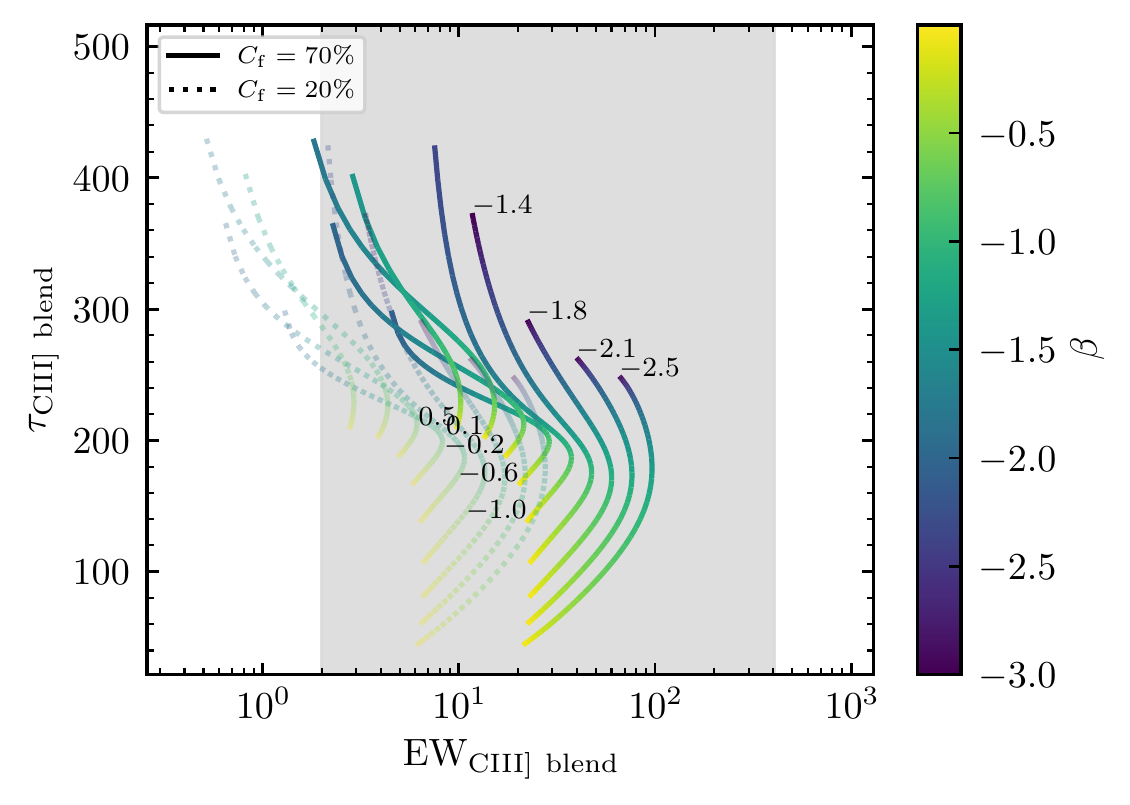} \\
    \includegraphics[width=0.48\textwidth]{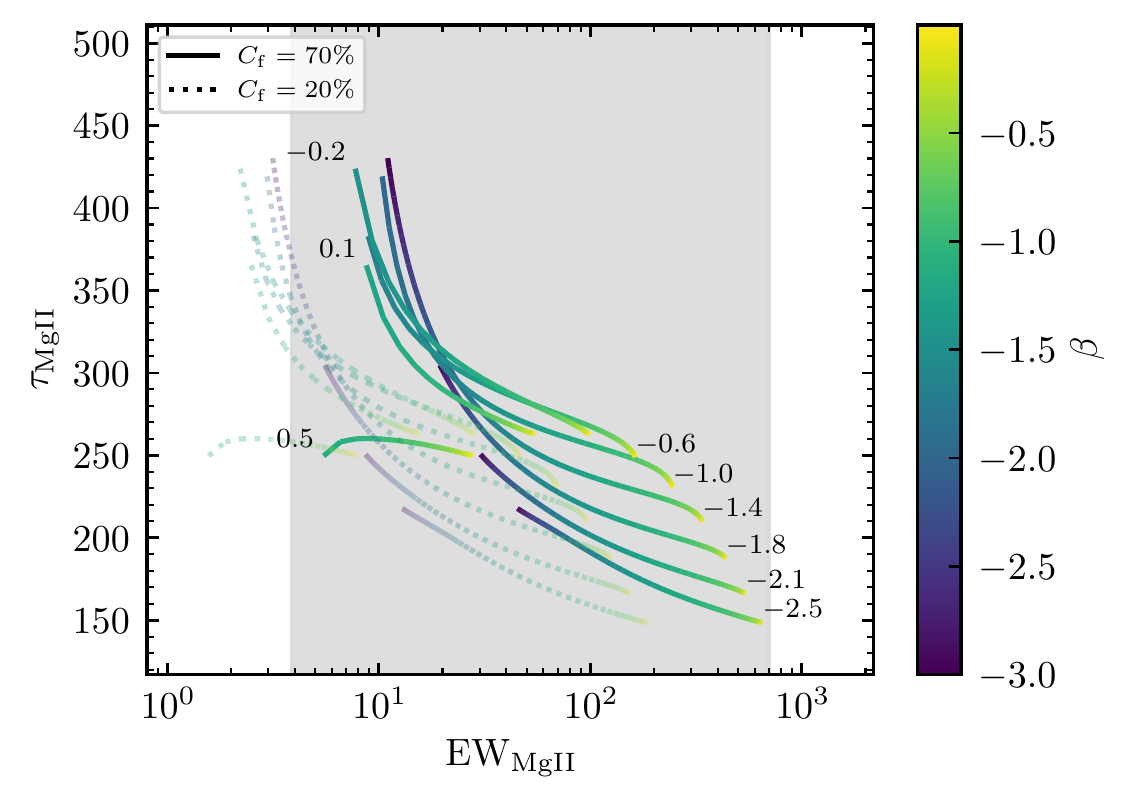}
    \includegraphics[width=0.48\textwidth]{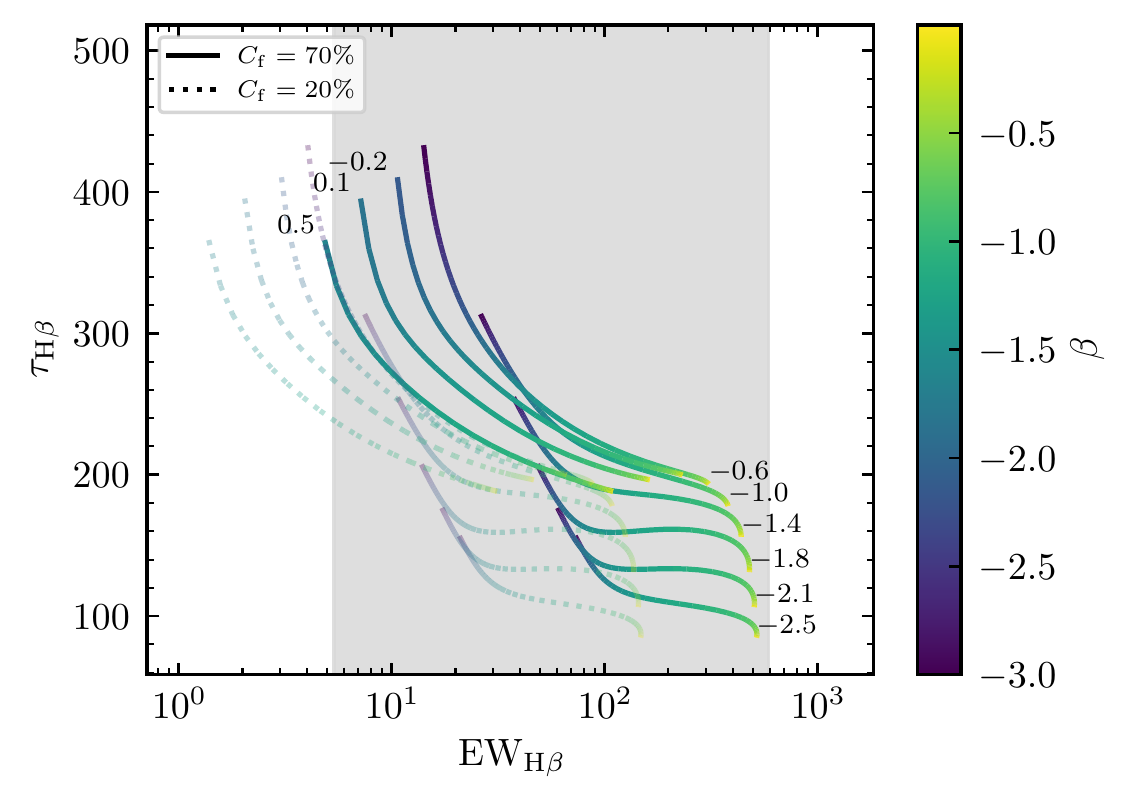}
    \caption{The correlations between EWs and time lags.
    The solid and dashed lines are the cases of $C_{\rm f}=70\%$ and
    $C_{\rm f}=20\%$, respectively. Each line has the same luminosity or
    Eddington ratio, and is color-coded by $\beta$. The Eddington ratio is
    marked nearby the corresponding line. The observations of the EW ranges of
    different emission lines (in Figure \ref{fig:ew_beta}) are also overlapped
    in grey. \label{fig:ew_lag}}
\end{figure*}

We calculate the transfer functions for different $\beta$ and show the results
in Figure \ref{fig:transfer_functions} for three cases of different Eddington
ratios [$\log (L_{\rm Bol}/L_{\rm Edd}) = -1.5$, $-0.5$, and $0.5$]. The mass of
SMBH is set to be $10^8 M_{\odot}$. Generally speaking, at low Eddington ratios,
the transfer functions are always positive and their peaks (with the strongest
responses) move toward longer time lags if $\beta$ decreases (more rarefied
BLRs). Along with Eddington ratio increases, the transfer functions with smaller
$\beta$ (more rarefied BLRs) still have longer time lags than those with larger
$\beta$ (denser ones), however their amplitudes become from positive to
negative. Negative transfer functions mean that the emission-line light curves
show inverse responses with respect to the variations of the continuum light
curves, which are different from the usual cases in RM. To quantify whether or
not the response is negative in average, we calculate the average of a transfer
function by $\bar{\Psi} = \int \Psi(t) dt / \int dt$. $\bar{\Psi}$ for different
$\beta$ are also shown in Figure \ref{fig:transfer_functions}. The transition
from positive $\bar{\Psi}$ to negative $\bar{\Psi}$ happens at larger $\beta$ if
Eddington ratio increases. More specifically, different lines show different
behaviors. At the same Eddington ratio, it is easier for the responses of Mg
{\sc ii}, C {\sc iv}, and C {\sc iii}] to become negative than those of
Ly$\alpha$, Si {\sc iv}$+$O {\sc iv}], and H$\beta$. The transitions of Mg {\sc
ii}, C {\sc iv}, and C {\sc iii}] occur at relatively larger $\beta$. For
example, Mg {\sc ii} can become negative at $\log (L_{\rm Bol}/L_{\rm Edd}) =
-0.5$, however the average responses of H$\beta$ keep positive in the same case.

The $\beta$ parameter controls both of EWs and typical BLR radii (time lags),
therefore we can investigate their correlations. We define an average time lag
of a transfer function using the formula of 
\begin{equation}
\tau=\int t \Psi(t) dt / \int \Psi(t) dt.
\label{eq:tau_mean}
\end{equation}
Through adjusting $\beta$, we can show the correlations between time lags and
EWs in Figure \ref{fig:ew_lag}. It should be noted that this definition is not
appropriate if $\Psi(t)$ is partly negative. Considering that currently only 
positive transfer functions are thought to be ``successful'' in RM observations,
we plot the cases of purely positive transfer functions here. Therefore,
the curves for high Eddington ratios in Figure \ref{fig:ew_lag} are cut off at
small $\beta$ (see, e.g., the cases of C {\sc iv} for $\log (L_{\rm Bol}/L_{\rm
Edd}) = 0.1$ and $0.5$ in Figure \ref{fig:ew_lag}).

\begin{figure*}
    \centering
    \includegraphics[width=0.48\textwidth]{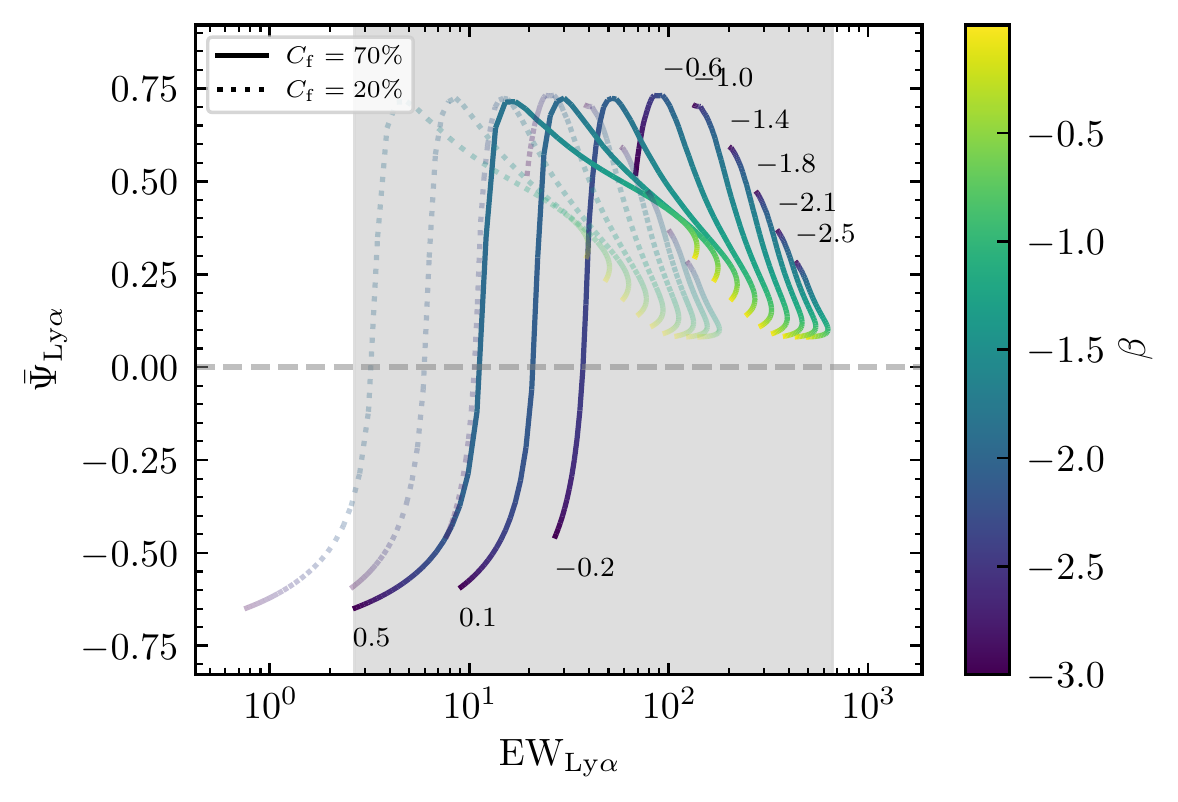}
    \includegraphics[width=0.48\textwidth]{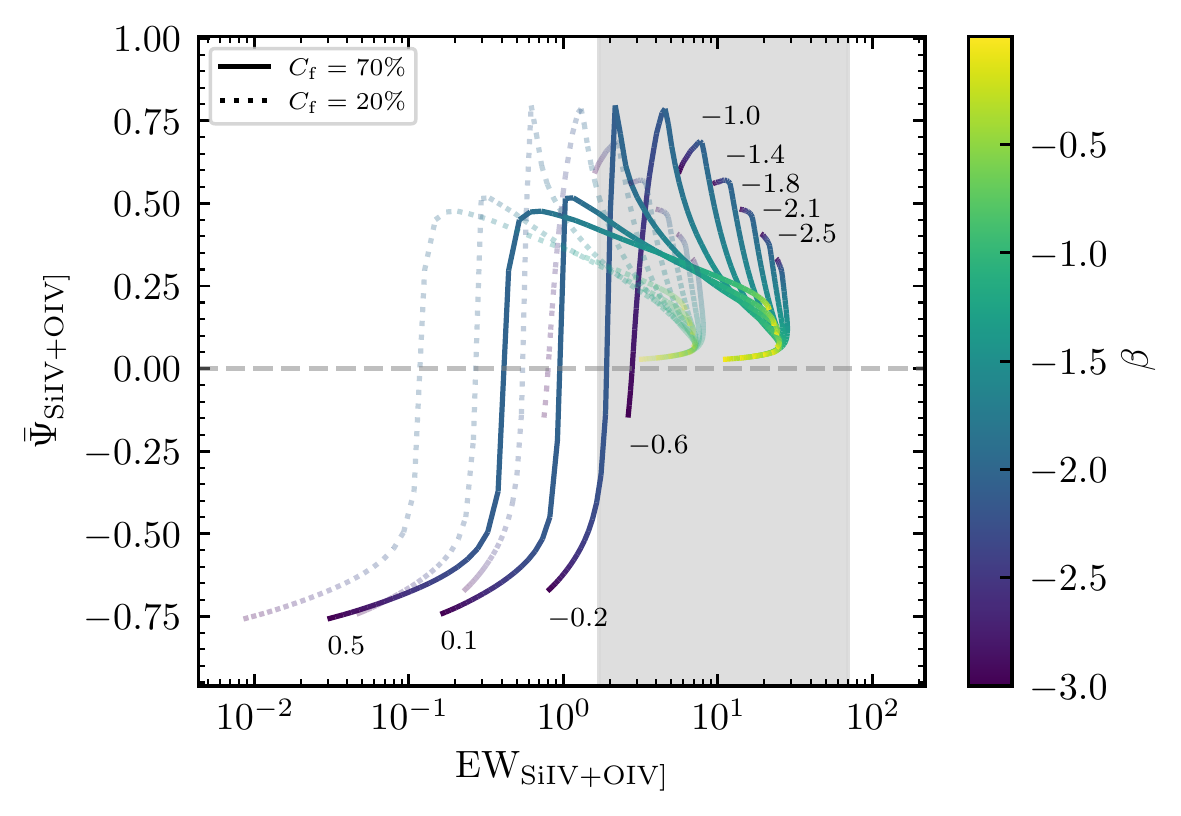} \\
    \includegraphics[width=0.48\textwidth]{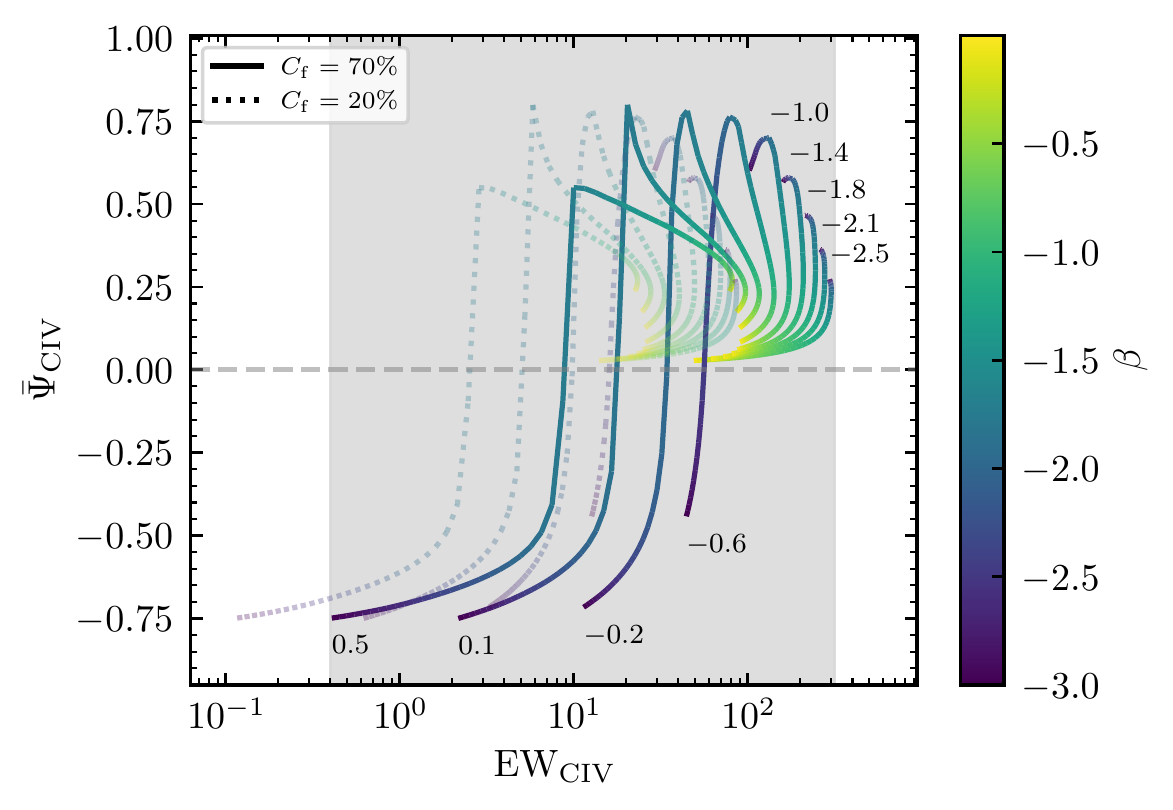}
    \includegraphics[width=0.48\textwidth]{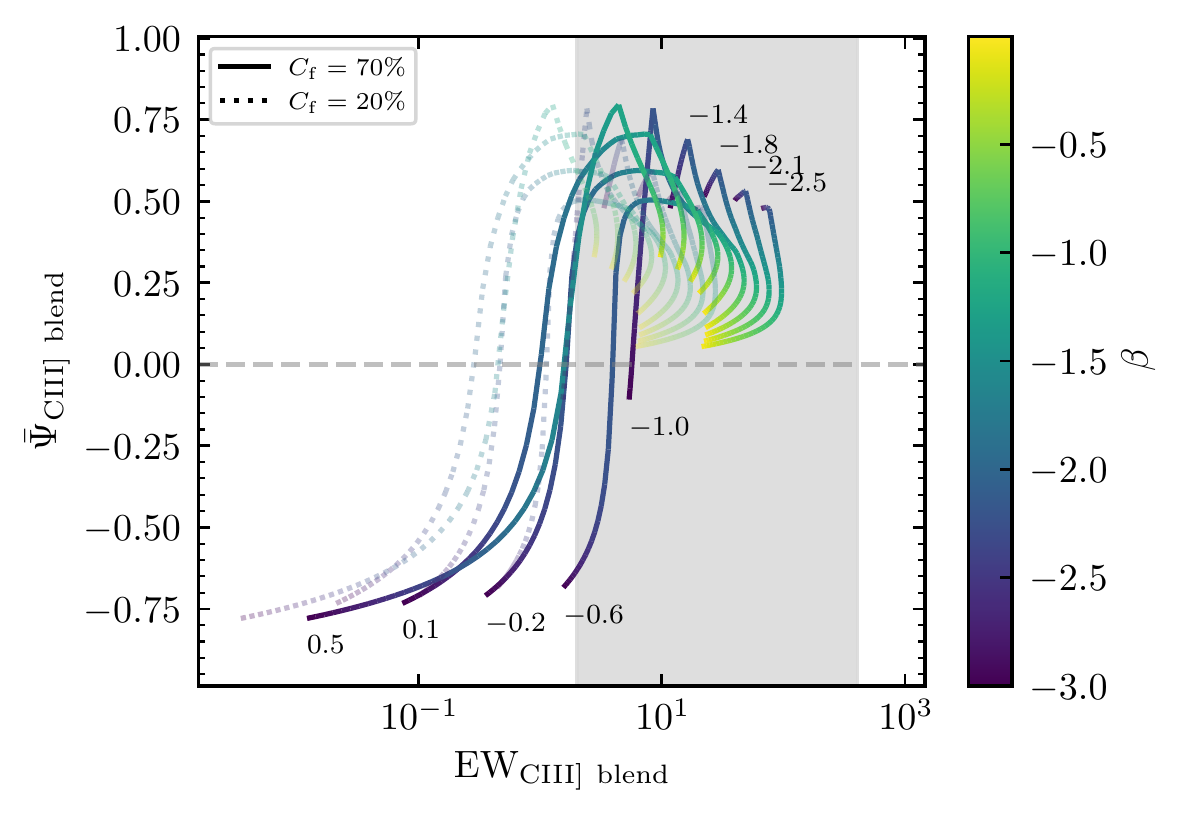} \\
    \includegraphics[width=0.48\textwidth]{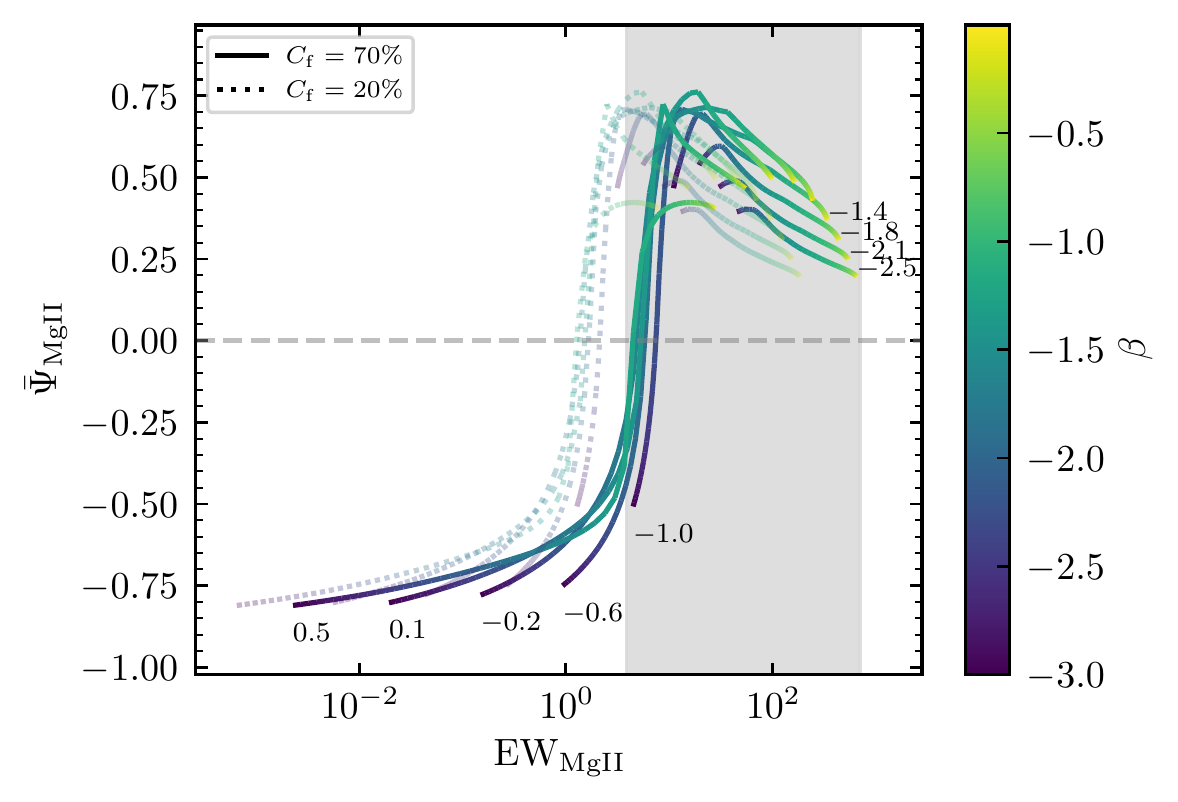}
    \includegraphics[width=0.48\textwidth]{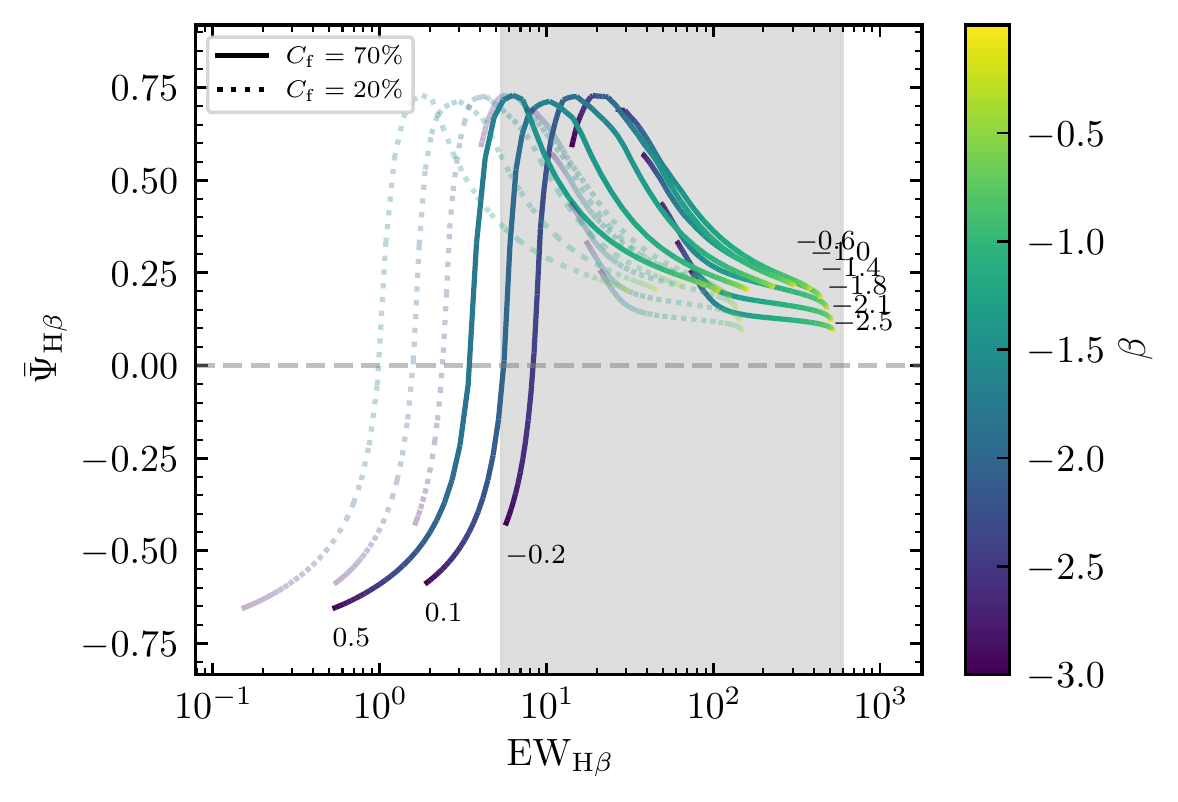}
    \caption{The correlations between EWs and the amplitudes of transfer
    functions. The solid and dashed lines are the cases of $C_{\rm f}=70\%$ and
    $C_{\rm f}=20\%$, respectively. Each line has the same luminosity or
    Eddington ratio, and is color-coded by $\beta$. The Eddington ratio is
    marked nearby the corresponding line. The observations of the EW ranges of
    different emission lines (in Figure \ref{fig:ew_beta}) are also overlapped
    in grey. \label{fig:ew_amplitude}}
\end{figure*}

In Figure \ref{fig:ew_lag}, each line has the same luminosity or Eddington ratio,
and is color-coded by the parameter $\beta$. These correlations mean that the
diverse density distributions of BLRs can result in variations of EWs (by nearly
an order of magnitude) and time lags (by factors of $\sim$2-3) simultaneously
even if the luminosities are the same. This may contribute to the scatters
($\sim$0.3dex) of the $R$-$L$ relationships of the emission lines and may also
be one of the explanations for the variations of time lags in individual objects
in different years. However, the correlations between time lags and EWs are
different in different emission lines. Ly$\alpha$, Si {\sc iv}$+$O {\sc iv}], C
{\sc iv}, and C {\sc iii}] show nonmonotonic correlations. In these four
emission lines, along with $\beta$ increasing, the time lags continuously become
shorter, however the EWs first increase and then decrease after across their
maximums. For the other two lines (Mg {\sc ii} and H$\beta$), both of the time
lags and EWs show almost monotonic correlations with $\beta$. $\beta$ shows
correlations (and anti-correlations) with their EWs (and time lags). More
specifically, for the three primary emission lines in RM - C {\sc iv}, Mg {\sc
ii}, and H$\beta$, they have the following typical characteristics.

C {\sc iv}: The EWs and time lags mainly show positive correlations at low
Eddington ratios [e.g., $\log (L_{\rm Bol}/L_{\rm Edd}) = -2.5$ and $-2.1$] and
anti-correlations at high Eddington ratios [e.g., $\log (L_{\rm Bol}/L_{\rm
Edd}) = 0.1$ and $0.5$]. The change of BLR density ($\beta$) causes the
variation of time lag but only weakly influence the EWs at moderate Eddington
ratios [e.g., $\log (L_{\rm Bol}/L_{\rm Edd}) = -0.6$].

Mg {\sc ii}: The EWs and time lags are mainly anti-correlated.
Time lag decreases if EW increases at almost all Eddington ratios. However,
the slopes are steeper at low EWs than those at high EWs. 

H$\beta$: The behavior of H$\beta$ line is more similar to that of
Mg {\sc ii} line. However, at high EWs (e.g., ${\rm EW}\gtrsim80$-$100$ for the
cases of $C_{\rm f}=70\%$ and ${\rm EW}\gtrsim30$-$40$ for the cases of $C_{\rm
f}=20\%$), the variations of the time lags are very weak along with the change
of $\beta$.

Here we do not plot the RM observations in Figure \ref{fig:ew_lag} because the
exact values of time lags are also significantly controlled by BLR geometry,
which is simply assumed to be spherically symmetric in our calculations (may be
very different from the actual situations). But we can generally do a simple
comparison. \cite{Du2019} shows the correlation between $\Delta R_{\rm H\beta} =
\log\,(R_{\rm H\beta}/R_{\rm H\beta,R-L})$ and ${\rm EW_{H\beta}}$, where
$R_{\rm H\beta,R-L}$ is the H$\beta$ lag calculated from the $R$-$L$
relationship. Each line in Figure \ref{fig:ew_lag} has the same luminosity.
Therefore, we can compare the observed $\Delta R_{\rm H\beta}$-${\rm
EW_{H\beta}}$ correlation with our calculations. In \cite{Du2019}, except for
those super-Eddington AGNs, the AGNs with normal accretion rates (with $70<{\rm
EW_{H\beta}}<300$) do not show any significant $\Delta R_{\rm H\beta}$-${\rm
EW_{H\beta}}$ correlation or anti-correlation. This is generally consistent with
our calculations that ${\rm EW_{H\beta}}$ and $\tau_{\rm H\beta}$ do not show
strong correlation at relatively high EWs. From our calculations, we expect that
more high-quality observations of the H$\beta$ with weaker EWs in AGNs with
normal accretion rates in future may discover some objects with longer time lags
than the $R$-$L$ relationship. 

In addition, we simply check the RM samples of C {\sc iv} lines \citep{Lira2018,
Grier2019, Kaspi2021} and obtain a very weakly positive $\Delta R_{\rm
CIV}$-${\rm EW_{CIV}}$ correlation with Spearman's rank correlation coefficient
of $\rho=0.32$ and a corresponding $p$-value of $0.09$ based on the C {\sc iv}
$R$-$L$ relationship in \cite{Kaspi2021}. Considering that the $\beta$ values of
the current C {\sc iv} RM samples are relatively large (see Figure
\ref{fig:ew_beta}), this may probably be consistent with our calculations,
especially the parts with $\tau<200$ days in Figure \ref{fig:ew_lag}. Similarly,
we also perform a test to the Mg {\sc ii} samples \citep{Lira2018,
Homayouni2020, Yu2021} based on the $R$-$L$ relationship in
\cite{Homayouni2020}. However, no significant correlation is found ($\rho=0.09$
and $p=0.61$). One possible reason is the relatively narrow EW span (0.16dex) of
the current Mg {\sc ii} RM samples \citep{Lira2018, Homayouni2020, Yu2021}. As a
comparison, the EW span of C {\sc iv} RM samples \citep{Lira2018, Grier2019,
Kaspi2021} is 0.30dex.

Through adjusting $\beta$, we can also plot the correlation between $\bar{\Psi}$
and EW in Figure \ref{fig:ew_amplitude} in order to check the EW at which the
transition from averagely positive to negative response happens. From Figure
\ref{fig:ew_amplitude}, C {\sc iv}, Ly$\alpha$ are more easier to show negative
(anomalous) responses in RM because the grey ranges cover more cases with
negative $\bar{\Psi}$. The observations of the other four lines can also cover
the negative responses from the photoionization calculations, however only if
the EWs are very close to the lower limits.

\begin{figure*}
    \centering
    \includegraphics[width=\textwidth]{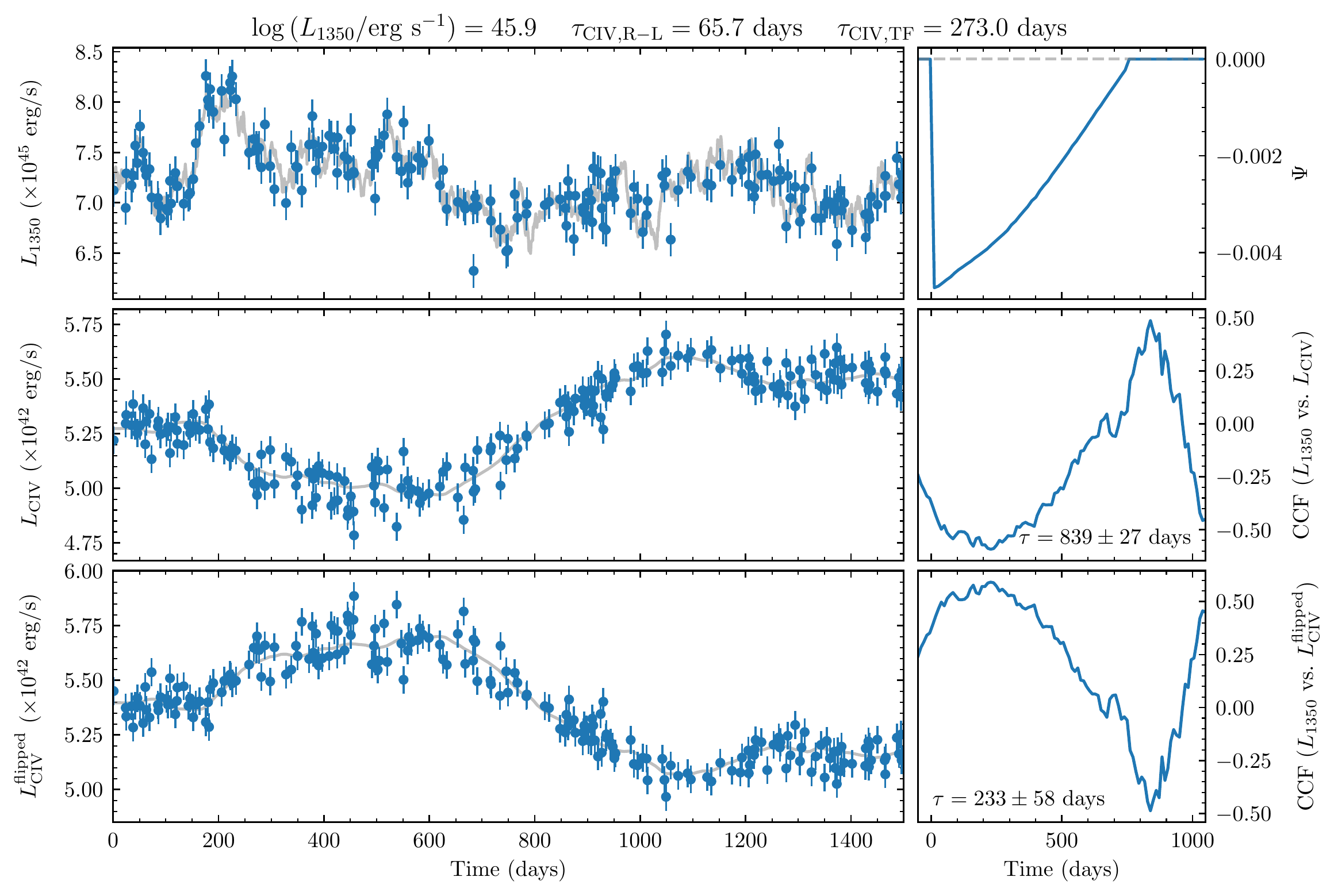}
    \caption{An example of C {\sc iv} light curve if $\beta=-2.0$. The
    grey lines are the mock light curves, and the blue points are the mock data
    by adding some artificial uncertainties ($10\%$) with a sampling cadence of
    10 days. The upper-left panel ($L_{1350}$) is the continuum light curve
    (from simple DRW model) in 1350\AA. The middle-left panel ($L_{\rm CIV}$) is
    the C {\sc iv} light curve. The lower-left panel ($L_{\rm CIV}^{\rm
    flipped}$) is the flipped C {\sc iv} light curve. The upper-right panel
    shows the transfer function $\Psi$. The middle-right panel is the CCF
    obtained from $L_{1350}$ vs. $L_{\rm CIV}$, which is the conventional
    analysis in RM but gives a wrong time lag ($\tau=839\pm27$ days) with
    respect to the input value ($\tau_{\rm CF,TF}=273.0$ days). The CCF from
    $L_{1350}$ vs. $L_{\rm CIV}^{\rm flipped}$ is provided in the lower-right
    panel and can yield the correct time lag ($\tau=233\pm58$ days). The input
    luminosity and the time lag from the transfer function are shown on the top.
    $\Psi$ is in arbitrary unit.
    \label{fig:light_curve}}
\end{figure*}

\section{Discussion}
\label{sec:discussion}

\subsection{Implication for RM observations}
\label{sec:ImplicationForRM}

As mentioned in Section \ref{sec:intro}, there are already several objects
showing anomalous behaviors in RM observations (especially CT~320, CT~803, and
2QZ~J224743 in \citealt{Lira2018}). The Si {\sc iv} line of CT~320 and the
Ly$\alpha$ line of CT~803 and 2QZ~J224743 show obviously negative responses (see
their light curves in Figure 5 of \citealt{Lira2018}). Their emission-line light
curves are inversely correlated with the continuum variations. The corresponding
cross correlations (CCFs) in \cite{Lira2018} exhibit strong troughs with minimum
cross-correlation coefficients smaller than $-0.5$. Positive peaks with such
amplitudes ($0.5$) in CCFs commonly represent significant responses. These
anomalous RM responses may probably be explained by their low BLR densities. 

In consideration of the possible negative responses of emission lines in
rarefied BLRs, it will lead to some troubles if their RM data are analyzed
according to traditional experiences. We demonstrate an example of the C {\sc
iv} negative response in Figure \ref{fig:light_curve}. It is obvious that the
emission-line light curve inversely responses the continuum light curve. In this
case, we adopt $\beta=-2.0$ and $\log (L_{\rm Bol}/L_{\rm Edd})=0.5$. For
simplicity, the continuum light curve is assumed to be a damped random walk
\cite[e.g.,][]{Zu2013}, which applies to most of AGNs
\cite[e.g.,][]{Kasliwal2015}. The monochromatic luminosity is $\log
(L_{1350}/{\rm erg\ s^{-1}}) = 45.9$, and the corresponding time lag is
$\tau_{\rm CIV, R-L}=65.7$ days based on the C {\sc iv} R-L relationship in
\cite{Kaspi2021}. The transfer function $\Psi$ is totally negative in this case.
The mean time lag calculated from transfer function is $\tau_{\rm CIV, TF} =
273.0$ days (Eqn \ref{eq:tau_mean}). To simulate observed light curves, some
artificial error bars (10\% for both continuum and emission line) are added. 

Conventionally, RM measures the time lag between the continuum and emission-line
light curves using CCF in order to determine the mass of SMBH. However, in this
simple example, we obtain a very different time lag of $\tau=839\pm27$ days
(directly from the strongest positive peak in the CCF) comparing with the above
input value ($\tau_{\rm CIV, TF}$) if we perform the time-series analysis
directly to the mock light curves using interpolated CCF \citep{Gaskell1987}.
The error bar of the time lag is obtained by the ``flux randomization/random
subset sampling'' method \citep[e.g.,][]{Peterson1998, Peterson2004}. This time
lag is obviously incorrect. The low peak correlation coefficient ($<0.5$) also
indicates that the continuum and emission-line light curves are poorly
correlated. 

If we flip the emission-line light curve, we can get a reliable time lag of
$\tau=233\pm58$ days, which is consistent with the input $\tau_{\rm CIV, TF}$
within $1\sigma$ uncertainties. The peak correlation coefficient between the
continuum and the flipped emission-line light curves is much higher (close to
0.7). Therefore, we need to flip the emission-line light curves in the
time-series analysis for such rarefied BLRs if we hope to get reliable time
lags. 

In practice, the maximum and minimum correlation coefficients
in CCF can be used as a criterion to identify the rarefied BLRs in real data. If
the absolute value of the minimum correlation coefficient is significantly
larger than that of the maximum correlation coefficient in an object with very
small emission-line EW and high Eddington ratio, it is probably a source with
rarefied BLR and should be analyzed by flipping its emission-line light curve.

\subsection{Justify Anemic BLR Model from RM Observations}
\label{sec:WLQ}

The physics behind WLQs has not been fully understood yet. Several models were
proposed to explain the origin of their emission-line weakness, and can
generally be divided into two categories. The first category is based on unusual
ionizing continuum (the deficit of ionizing photons from accretion disks) due
to, e.g.,  high accretion rates \citep{Leighly2007a, Leighly2007b}, the
absorption by some shielding gas \citep[][]{Wu2011}, the shielding by the
puffed-up inner region of the slim accretion disks \citep[][]{Luo2015}, or even
the very cold accretion disks in hyper-massive black holes \citep{Laor2011}. The
second category is the anemic BLR model in which the BLRs themselves are lack of
gas \citep[e.g.,][]{Shemmer2010}. The BLRs are probably in the very early stage
of formation \citep[][]{Hryniewicz2010, Wang2012, Andika2020}. A question which
arises is how can we further reveal what happens in WLQs?

From the above photoionization calculations, we predict that the anemic BLR may
lead to negative response of C {\sc iv} lines with respect to the continuum
variation in RM observations. We propose that RM can be used to verify the
anemic BLR model in WLQs. From Figure \ref{fig:ew_beta}, we can obtain a general
constraints to the $\beta$ parameters in WLQs. The $\beta$ parameters of C {\sc
iv} and H$\beta$ emission lines in WLQs have been constrained to be within the
ranges of [$-2.0$, $-1.0$] and [$-0.7$, $0.0$] for the case with $C_{\rm
f}=20\%$, and [$-2.5$, $-1.5$] and [$-1.2$, $-0.5$] for the case with $C_{\rm
f}=70\%$, respectively. The response behaviors of these two lines are different.
The C {\sc iv} lines in WLQs can have negative $\bar{\Psi}$ (especially for
those with $-2.5 < \beta < -1.75$). However, from the photoionization
calculations and the range of $\beta$, the H$\beta$ emission lines in WLQs do
not have similar behaviors. The H$\beta$ lines of WLQs (at least the current WLQ
samples) only positively response to the variations of the continuum radiation. 

Therefore, from the negative response of C {\sc iv} emission line to the
variation of the continuum, it is possible to validate the anemic BLR model of
the WLQs. If some WLQs are found to show negative C {\sc iv} response in RM
observations, it implicates that the anemic BLR model works (at least in some
WLQs). On the contrary, if none of the C {\sc iv} in WLQs exhibits any negative
response, the anemic model doesn't work and the model based on unusual ionizing
continuum may play a key role in WLQs. 

Because of the weakness of the emission lines in WLQs, high-fidelity RM
observations with highly-accurate flux calibration from large-aperture
telescopes are required (e.g., Gemini 8.1 m, Magellan 6.5 m telescopes).
Considering that the typical EWs of C {\sc iv} in WLQs are lower than their
normal counterparts by roughly factors of 5$\sim$10, the accuracy of flux
calibration should be as good as $0.5\%\sim1\%$ (given that the typical error
bars of the current C {\sc iv} RM observations are $4 \sim 5\%$, see, e.g.,
\citealt{Lira2018, Kaspi2021}). There is no any narrow emission lines in the UV
spectra of WLQs in their rest frames, the traditional narrow-line-based
calibration method in RM \citep[e.g.,][]{Peterson1998, Bentz2009, Grier2017}
cannot be performed in WLQs. Instead, the comparison-star-based calibration
\citep[][]{Kaspi2000, Kaspi2021, Du2014} should be adopted. This method can in
principle provide good calibration accuracy ($\sim 1\%$) in relatively good
weather condition. 

Using the latest R-L relation \citep[e.g.,][]{Lira2018, Hoormann2019, Grier2019,
Kaspi2021}, the time lags of C {\sc iv} can be estimated if the monochromatic
luminosities are given. However, the phenomenon of shortened time lags in high
accretion rate AGNs found in H$\beta$ emission lines \cite[e.g.,][]{Du2015,
Du2016, Du2018} may also play roles in C {\sc iv} lines \citep{DallaBonta2020}.
WLQs have relatively high accretion rates \citep[][see also Figure
\ref{fig:ew_beta}]{Leighly2007a, Leighly2007b, Luo2015}. Therefore, the sampling
cadences for WLQs need to be higher than the expected from the C {\sc iv} R-L
relation.

It should be noted that the shielding gas model of WLQs may
also lead to anomaly in the emission-line responses, similar to the cases of
``BLR holiday'' in NGC~5548 \citep{Goad2016, Pei2017} and Mrk~817
\citep{Kara2021} that the continuum and emission-line light curves are
decoupled. The changes of the properties (density, covering factor, etc.) of the
shielding gas \citep[or disk wind,][]{Wu2011, Luo2015, Jin2022} can mainly
influence the emission line but do not significantly affect the continuum if the
shielding gas is not in the line of sight. In this case, the continuum and
emission-line light curves may probably show weak (or even no) correlations. This
kind of anomaly is different from the negative responses discussed in the
present paper.

\subsection{Possible Explanation for Weak C {\sc iv} and Normal H$\beta$:
Radiation Pressure}
\label{sec:WLQ_Hb_CIV}

A puzzle in WLQs is why high ionization lines (e.g., C {\sc iv}) show relatively
weak EWs however low ionization lines (like H$\beta$) do not. One possibility is
that the ionizing continuum for the BLRs in WLQs is unusually soft due to
super-Eddington accretion \citep{Leighly2007a,Leighly2007b}, shielding gas
\citep{Wu2011}, the central puffed-up inner regions of slim disks
\citep{Luo2015}, or the code accretion disks in hyper-massive AGNs
\citep{Laor2011}. The other possibility is that the gas clouds of high- and
low-ionization lines have different physical properties in ``anemic'' model
\citep{Plotkin2015}. In the context of our photoionization calculation, the
observations of C {\sc iv} and H$\beta$ EWs indicate that only C {\sc iv}
emitting clouds suffer ``anemia'' however the H$\beta$ clouds tend to be more
``normal''. Only $\beta_{\rm CIV}$ becomes significantly smaller than $-1$ in
WLQs, but $\beta_{\rm H\beta}$ is still close to $-1$ (see Figure
\ref{fig:ew_beta} and Section \ref{sec:WLQ}).

It has been known for many years that the C {\sc iv} lines in AGNs tend to show
blueshifted profiles which are usually interpreted by outflow kinematics
\citep[e.g.,][]{Marziani1996, Baskin2005, Richards2011}. A clear demonstration
of the C {\sc iv} outflow came from the RM observation of NGC~5548 in UV band
\citep{Bottorff1997}. On the contrary, the outflow kinematics is relatively rare
in the H$\beta$ emitting region. The velocity-resolved RM reveals
that the H$\beta$ kinematics of most AGNs are dominated by virialized motion or
inflow \citep[e.g.,][or see Figure 14 in \citealt{U2022}]{Bentz2010, Grier2013,
DeRosa2018, Bao2022}. The high-quality two-dimensional transfer function of H$\beta$ in
NGC~5548 shows that its H$\beta$ region is a Keplerian rotating disk
\citep{Xiao2018, Horne2021}. 

\begin{figure}[!t]
    \centering
    \includegraphics[width=0.5\textwidth]{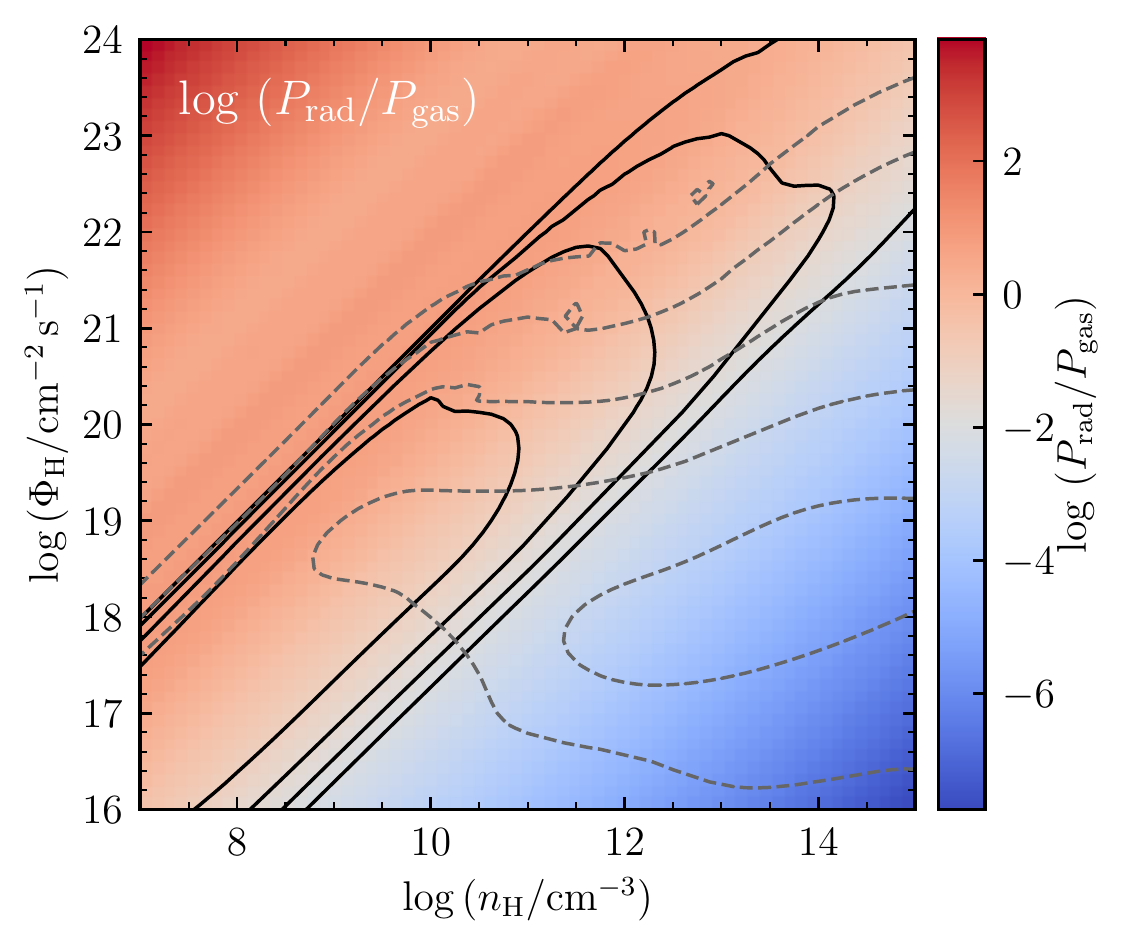}
    \caption{Ratio between radiation pressure ($P_{\rm rad}$) and gas pressure
    ($P_{\rm gas}$). The EW contours of C {\sc iv} (solid lines) and H$\beta$
    (dashed lines) are overlayed. It is obvious that the radiation pressure on
    the C {\sc iv}-emitting clouds are much stronger than the H$\beta$-emitting
    clouds. \label{fig:pressure}}
\end{figure}

We check the radiation pressure $P_{\rm rad}$ (due to the attenuation of the
incident continuum) and gas pressure $P_{\rm gas}$ of the photoionization grid
in Section \ref{sec:photoionization}. The ratio $P_{\rm rad}/P_{\rm gas}$
overlayed with the contours of the C {\sc iv} and H$\beta$ EWs is shown in
Figure \ref{fig:pressure}. It is obvious that the radiation pressure acting on
the clouds is generally larger than the gas pressure in the C {\sc iv} emitting
region, but smaller in the H$\beta$ region. This may probably explain, at least
in the framework of anemic BLR model, why only C {\sc iv} lines become much
weaker in WLQs. The large radiation pressure on C {\sc iv} clouds may drive
strong outflow and push the medium away. This process undoubtedly reduces the
gas content (gas density). WLQs have generally higher Eddington ratios than
normal quasars and hence stronger radiation pressures (stronger C {\sc iv}
outflow). The radiation pressure on H$\beta$ clouds is much weaker, therefore
the gas is still bounded by the gravitational potential of the central SMBH.
This speculation is also consistent with the vertical geometry of BLRs in
\cite{Kollatschny2013} that C {\sc iv} clouds are relatively far away from the
mid-plane (blown away by radiation pressure) but H$\beta$ is emitted in a more
flattened geometry. 

The other UV lines aforementioned also have similar behaviors as C {\sc iv}. Si
{\sc iv}$+$O {\sc iv}] and C {\sc iii}] emitting clouds may suffer stronger
radiation pressure. The radiation pressure may be slightly weaker on Ly$\alpha$
clouds, and much weaker on Mg {\sc ii} clouds. Therefore, the WLQ phenomena and
anomalous responses on Si {\sc iv}$+$O {\sc iv}], C {\sc iii}], and Ly$\alpha$
may probably stronger than Mg {\sc ii}. 

\subsection{Other Factors: SED, metallicity, $\Gamma$, and
Covering Factor?}
\label{sec:other_factors}

We adopted the SED from \cite{Mathews1987} in our photoionization calculation.
However, SED depends on BH masses and Eddington ratios of AGNs \citep[e.g.,][and
references therein]{Jin2012, Ferland2020}. In addition, the BLR metallicity is
probably higher in the AGNs with high accretion rates \citep[e.g.,][]{Panda2019,
Sniegowska2021}. Considering that WLQs have high accretion rates
\citep[e.g.,][see also Figure \ref{fig:ew_beta}]{Shemmer2010, Wu2011,
Plotkin2015}, we adopt a different input configuration (with the SED for the
highest Eddington ratio in \cite{Ferland2020} and higher metallicity) and run
the calculation again as a test (see more details in Appendix \ref{sec:app}). A
smaller $\Gamma$ is also adopted here. We find that the general conclusions in
the present paper (e.g., EW vs. $\bar{\Psi}$ and negative response of C {\sc
iv}) do not change significantly (see Appendix \ref{sec:app}). More realistic
calculations are still needed in future.

Actually, the other possibility of the anemic BLR model is that WLQs have very
anomalous covering factor of their BLR clouds but similar gas densities as their
normal counterparts. If this is the case, their C {\sc iv} will not show
negative responses to the varying continuum.  

Note that here we assume the BLR geometry is spherically symmetric, which could
be too simple. The true BLRs may be thick disks or even more complex geometry or
kinematics (inflow or outflow, e.g., see \citealt{Pancoast2014, Williams2018,
Villafana2022}). Comparing with the spherically symmetric
cases in the present paper, the major difference of thick BLR disks is that they
have fewer gas clouds and/or lower gas density at the regions in the polar
directions and relatively far away from the central ionizing sources. Therefore,
it is expected that the transfer functions in thick BLR disks will be narrower
than the cases shown in the present paper, and/or the responses may become
negative at relatively larger $\beta$. However, the general tendency, that
rarefied BLRs (small $\beta$) show negative responses, will remain the same.
We will perform more sophisticated calculation for practical
BLR geometry and kinematics in future.

\section{Summary}
\label{sec:summary}

In this paper, we present photoionization calculations (LOC models) for the
one-dimensional transfer functions of Ly$\alpha$ $\lambda1216$, Si {\sc iv}$+$O
{\sc iv}] $\lambda1400$ blend, C {\sc iv} $\lambda1549$ doublet, C {\sc iii}]
$\lambda1909$ blend, Mg {\sc ii} $\lambda2798$ doublet, and H$\beta$
$\lambda4861$ emission lines in order to investigate the roles of BLR densities
in their RM observations. Based on the calculations and the comparison with 
observations, we have the following predictions and conclusions:

\begin{itemize}
    \item The AGNs with rarefied BLRs (small $\beta$) are predicted to show
    negative responses (anomalous responses) in RM observations. The emission
    lines of such objects may have relatively low EWs. The observed anomalous
    behaviors in the UV emission lines of some objects in the past RM campaigns
    (e.g., CT~320, CT~803, and 2QZ~J224743 in \citealt{Lira2018}) may be
    explained by the rarefied BLRs.  In this case, the emission-line light
    curves may need to be flipped before the time-series analysis if we want to
    get accurate BLR radii.

    \item The different BLR densities in AGNs may contribute to the scatter of
    the $R$-$L$ relationship. The wide distributions of the BLR densities can
    result in changes of time lags by factors of 2-3. Preliminarily, the
    observed scatter of the C {\sc iv} $R$-$L$ relationship is probably
    consistent with the calculations in the present paper. For the other
    emission lines (Mg {\sc ii} and H$\beta$), more RM observations for the AGNs
    with wider EW spans are needed.

    \item The variation of time lags in individual objects without significant
    changes of continuum luminosities may be explained by the changes of BLR
    densities with time. 

    \item We propose that the existence of negative responses in C {\sc iv} RM
    observations can be used to justify if the anemic BLR model works or not in
    WLQs. If negative responses are found in WLQs, their emission-line weakness
    can be attributed to the deficit of BLR gas.
\end{itemize}

\acknowledgments

We thank the anonymous referee for the useful comments that
improved the manuscript. We acknowledge the support by National Key R\&D
Program of China (grants 2021YFA1600404, 2016YFA0400701), by the National
Science Foundation of China through grants {NSFC-12022301, -11991051, -11991054,
-11873048, -11833008}, by Grant No. QYZDJ-SSW-SLH007 from the Key Research
Program of Frontier Sciences, Chinese Academy of Sciences (CAS), and by the
Strategic Priority Research Program of CAS grant No.XDB23010400.

\appendix

\section{A Test for Input Configuration}
\label{sec:app}

The input configuration (SED, BLR metallicity, and $\Gamma$) adopted in our
photoionization calculation of the main text is appropriate for typical quasars.
However, WLQs tend to have higher accretion rates \citep[e.g.,][see also Figure
\ref{fig:ew_beta}]{Shemmer2010, Wu2011, Plotkin2015}, thus this configuration
may be not ideal for WLQs. Here we change the configuration and run the
calculation again for comparison. We adopt the SED for the highest Eddington
ratio in \cite{Ferland2020} and a metallicity of $5Z_{\odot}$, where $Z_{\odot}$
is the solar abundance (the metallicity of high-accretion-rate quasars are
higher, see, e.g., \citealt{Panda2019, Sniegowska2021}). The BLR sizes of the
AGNs with high accretion rates measured from RM are smaller than the predictions
of the $R$-$L$ relation \citep[e.g.,][]{Du2015, Du2018}, which may possibly
imply that their BLR clouds are more concentrated in the central regions
(corresponding to smaller $\Gamma$). We thus adopt $\Gamma=-1.5$. The other
parameters are kept the same as the main text. We call this configuration
``Configuration B'' (The input for the photoionization calculation in the main
text is called ``Configuration A''). The results of Configuration B are shown in
Figures \ref{fig:ew_flux_eta_highaccr}-\ref{fig:ew_amplitude_highaccr}.

We find the general results do not change significantly. For example, if $\beta$
decreases, the transfer function tend to become negative, especially for the
cases with high Eddington ratios. The correlations between EWs and time lags may
contribute to the scatters of the $R$-$L$ relations. These are almost the same
as the results of Configuration A. But there are still some small differences
that we noticed. (1) The responsivities of Ly$\alpha$ and H$\beta$ in Figure
\ref{fig:ew_flux_eta_highaccr} have small positive zones in the high $\Phi_{\rm
H}$ and low $n_{\rm H}$ regions. (2) With the same Eddington ratio, the transfer
functions of Configuration B can become negative at larger $\beta$ (see Figure
\ref{fig:transfer_functions} and \ref{fig:transfer_functions_highaccr}). (3) The
correlations between EWs and time lags for Si {\sc iv}$+$O {\sc iv}, C {\sc iv},
and C {\sc iii} blend are more monotonic (see Figure \ref{fig:ew_lag_highaccr}),
which are a little different from the nonmonotonic correlations in Figure
\ref{fig:ew_lag}.

\begin{figure*}[!ht]
    \centering
    \includegraphics[width=\textwidth]{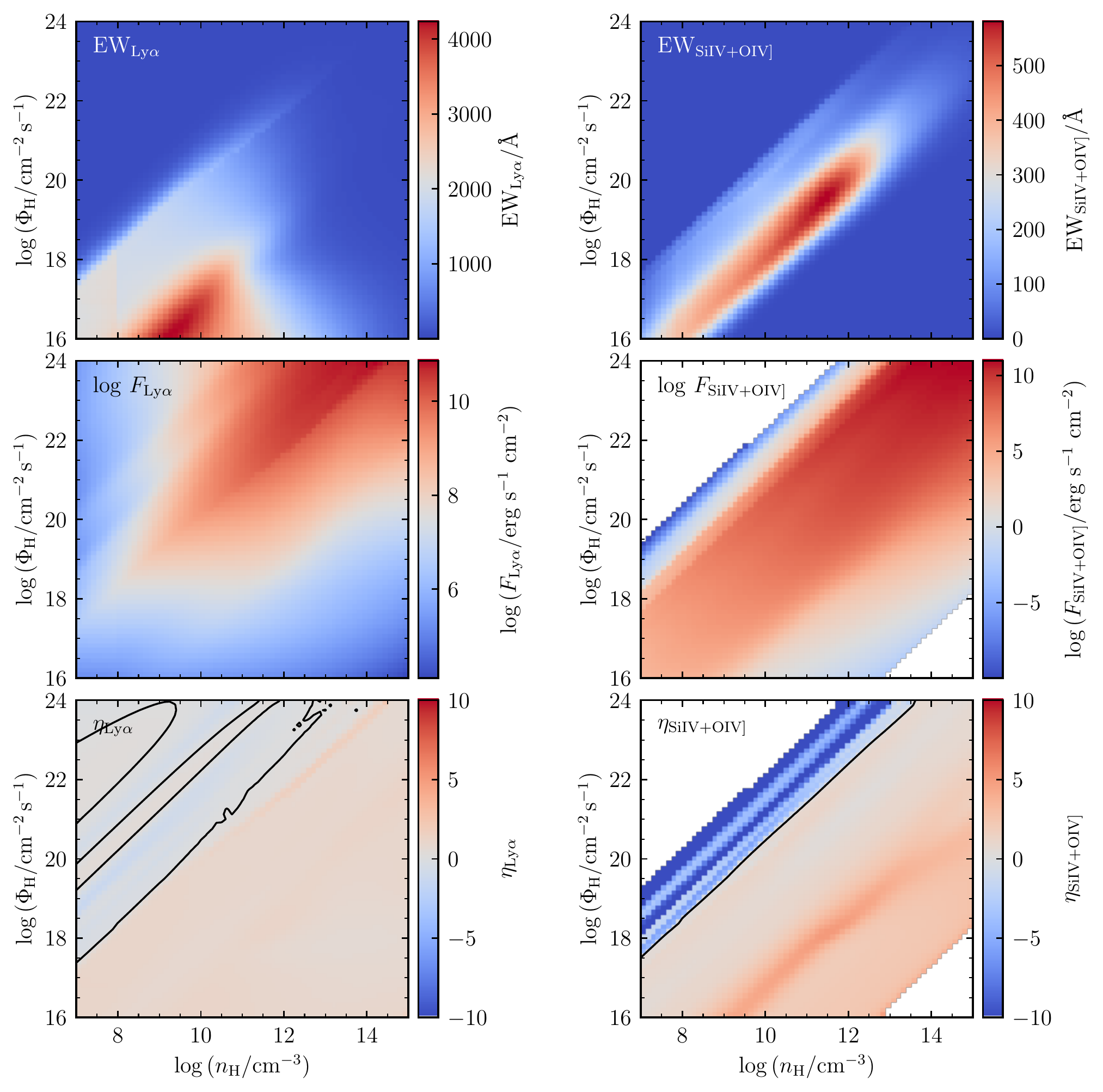}
    \caption{EWs, fluxes, and responsivity of the emission lines for different
    $n_{\rm H}$ and $\Phi_{\rm H}$ (Configuration B). The meanings of the
    panels, colors, and lines are the same as Figure \ref{fig:ew_flux_eta}.
    \label{fig:ew_flux_eta_highaccr}}
\end{figure*}

\begin{figure*}
    \figurenum{\ref{fig:ew_flux_eta_highaccr}}
    \centering
    \includegraphics[width=\textwidth]{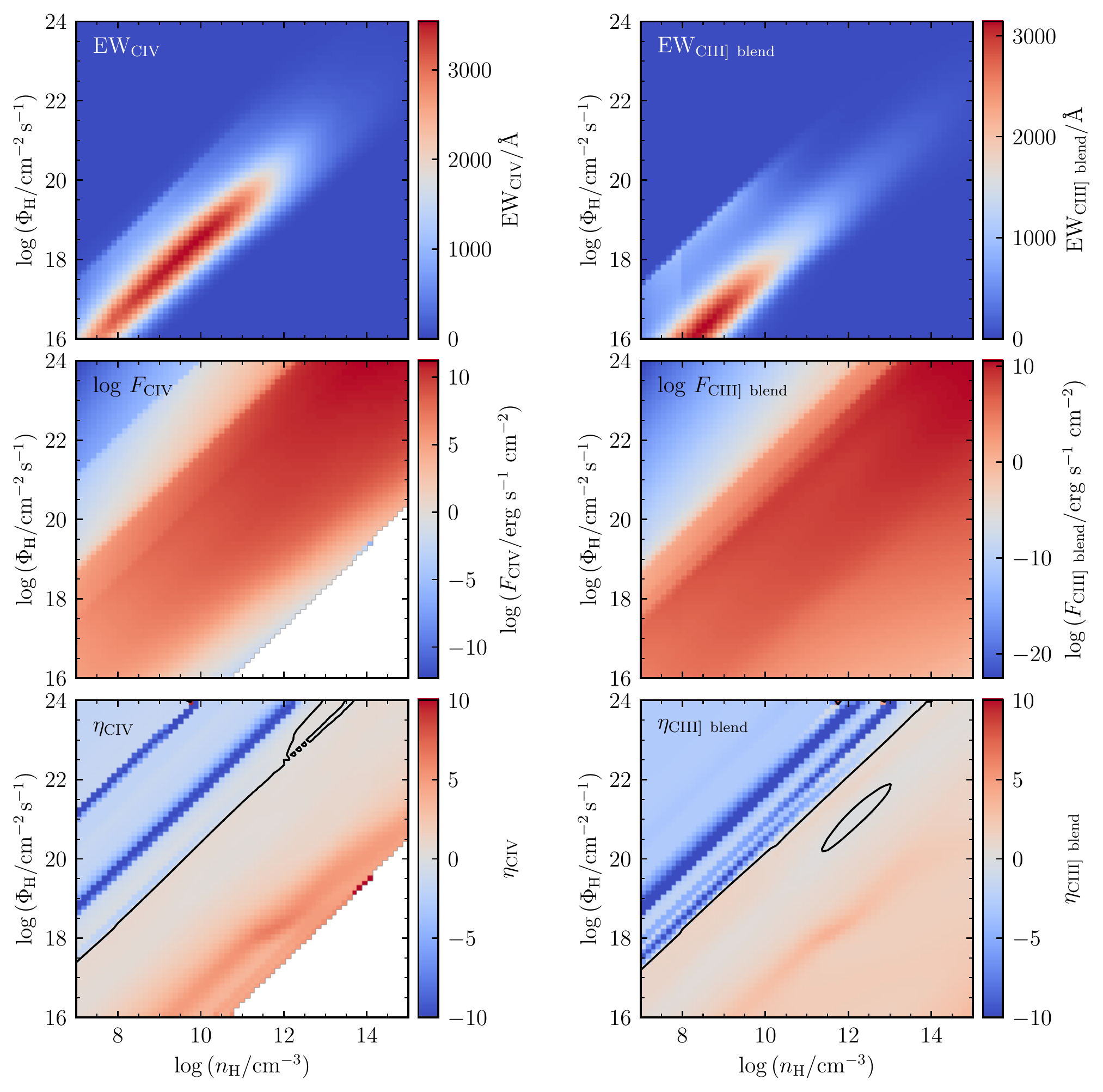}
    \caption{(Continued.)}
\end{figure*}

\begin{figure*}
    \figurenum{\ref{fig:ew_flux_eta_highaccr}}
    \centering
    \includegraphics[width=\textwidth]{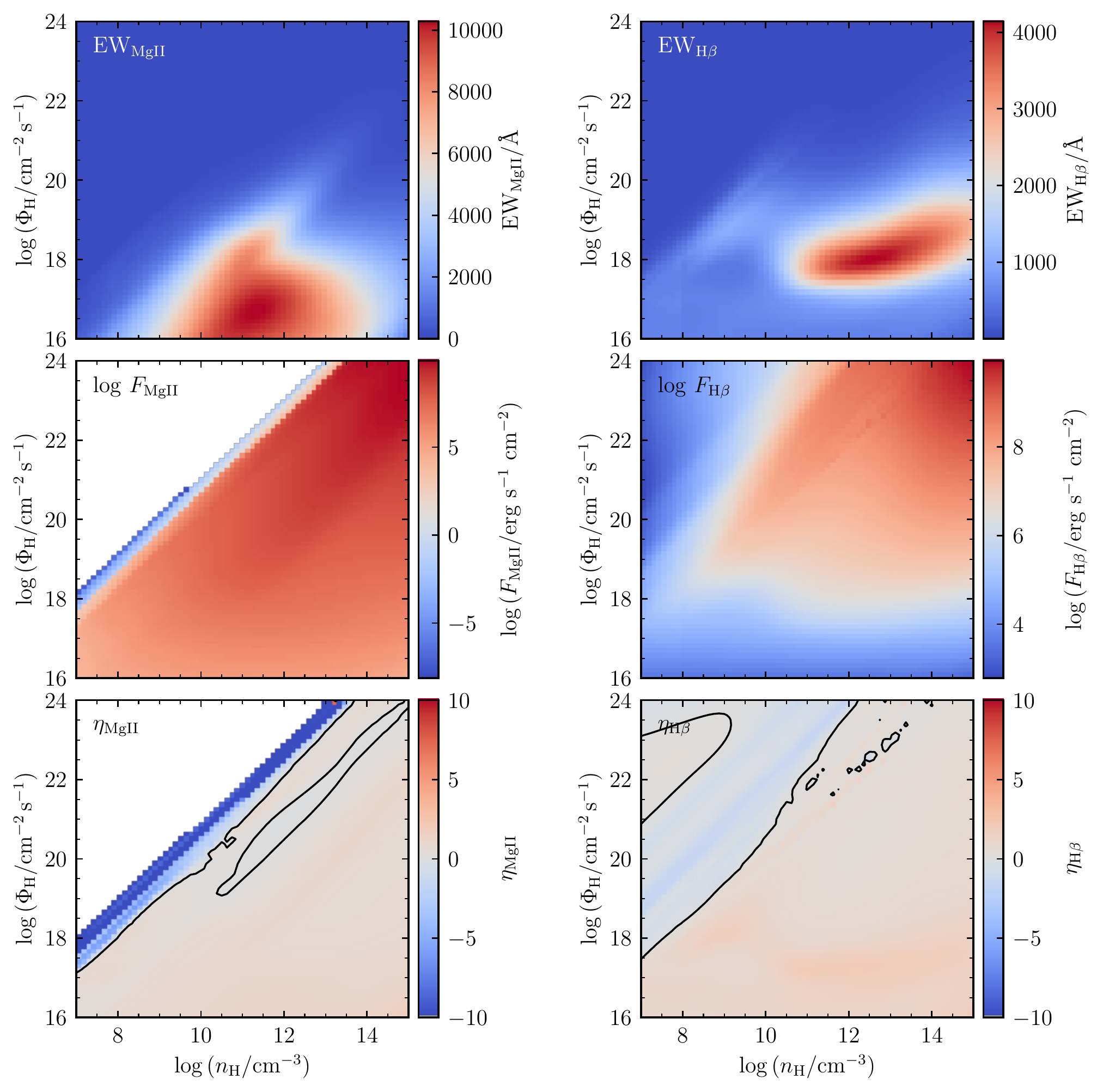}
    \caption{(Continued.)}
\end{figure*}

\begin{figure*}[!ht]
    \centering
    \includegraphics[width=\textwidth]{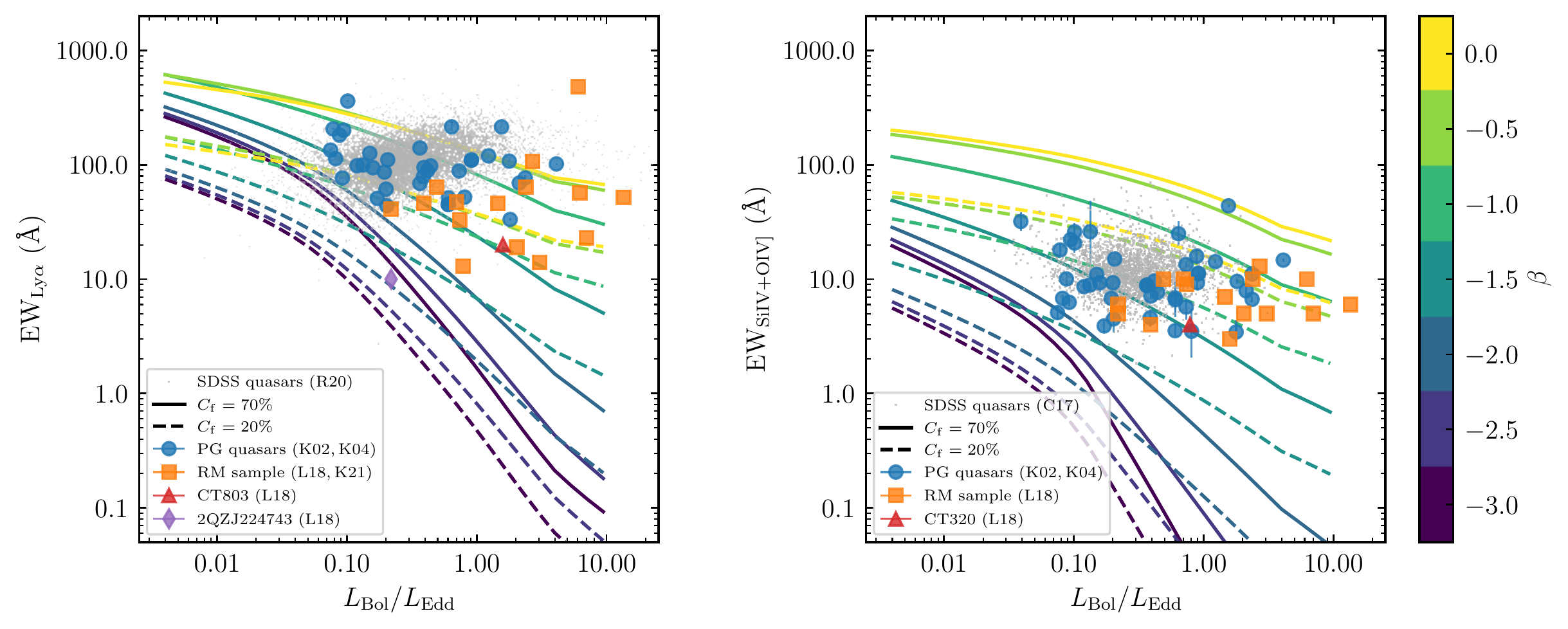}\\
    \includegraphics[width=\textwidth]{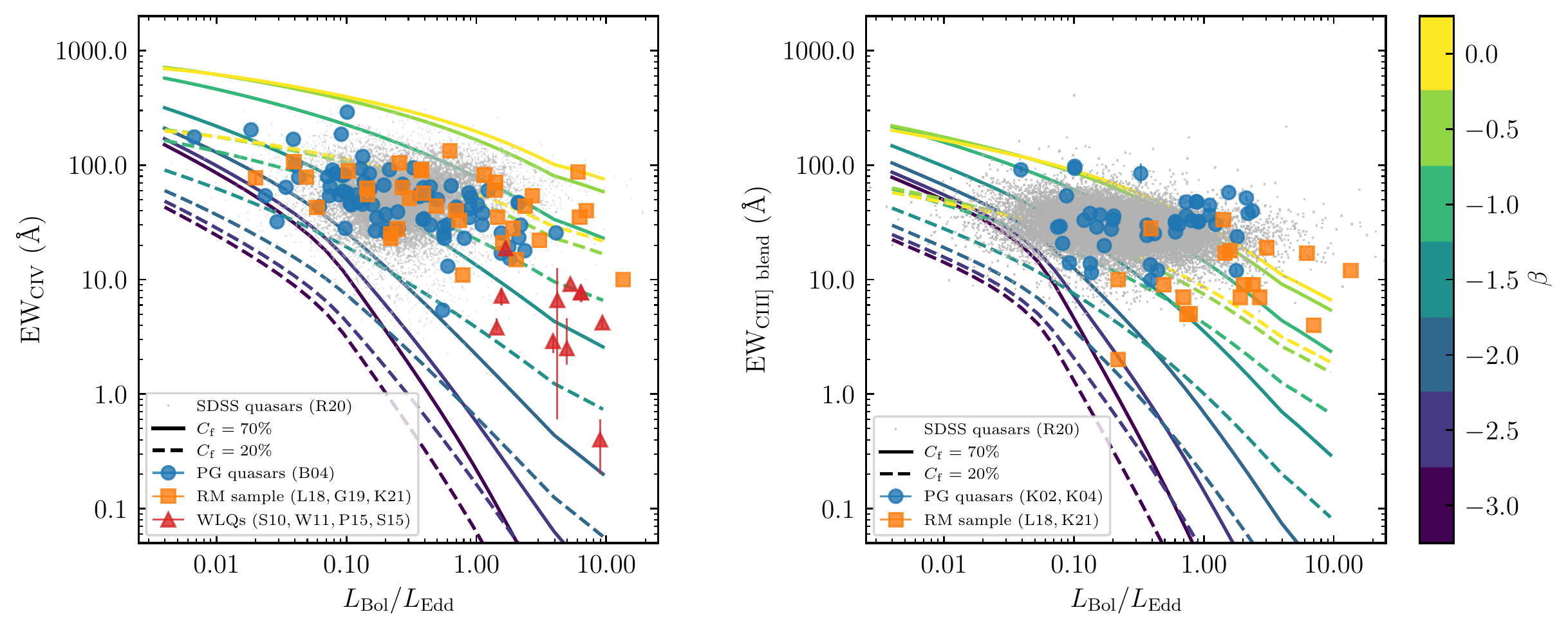}\\
    \includegraphics[width=\textwidth]{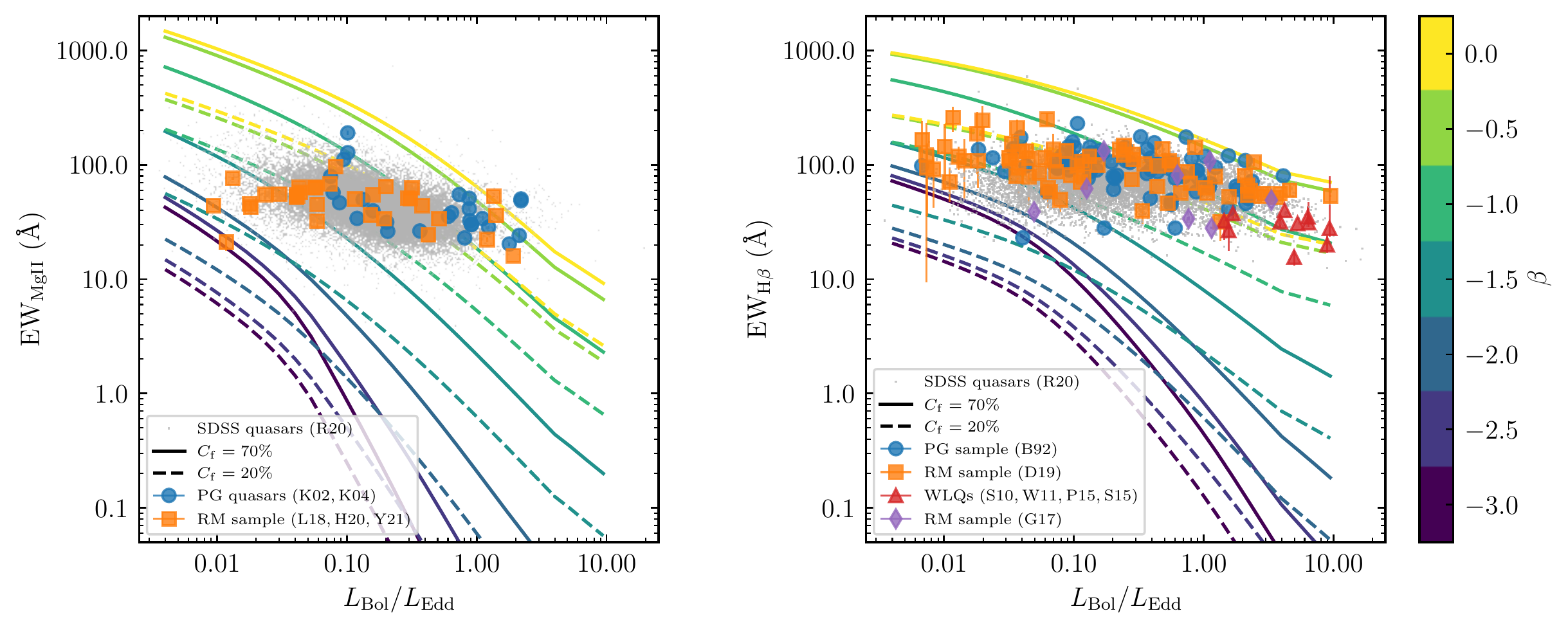}
    \caption{EW vs. Eddington ratio (Configuration B). The meanings of the
    panels, colors, lines, and symbols are the same as Figure \ref{fig:ew_beta}.
    \label{fig:ew_beta_highaccr}}
\end{figure*}

\begin{figure*}
    \centering
    \includegraphics[width=\textwidth]{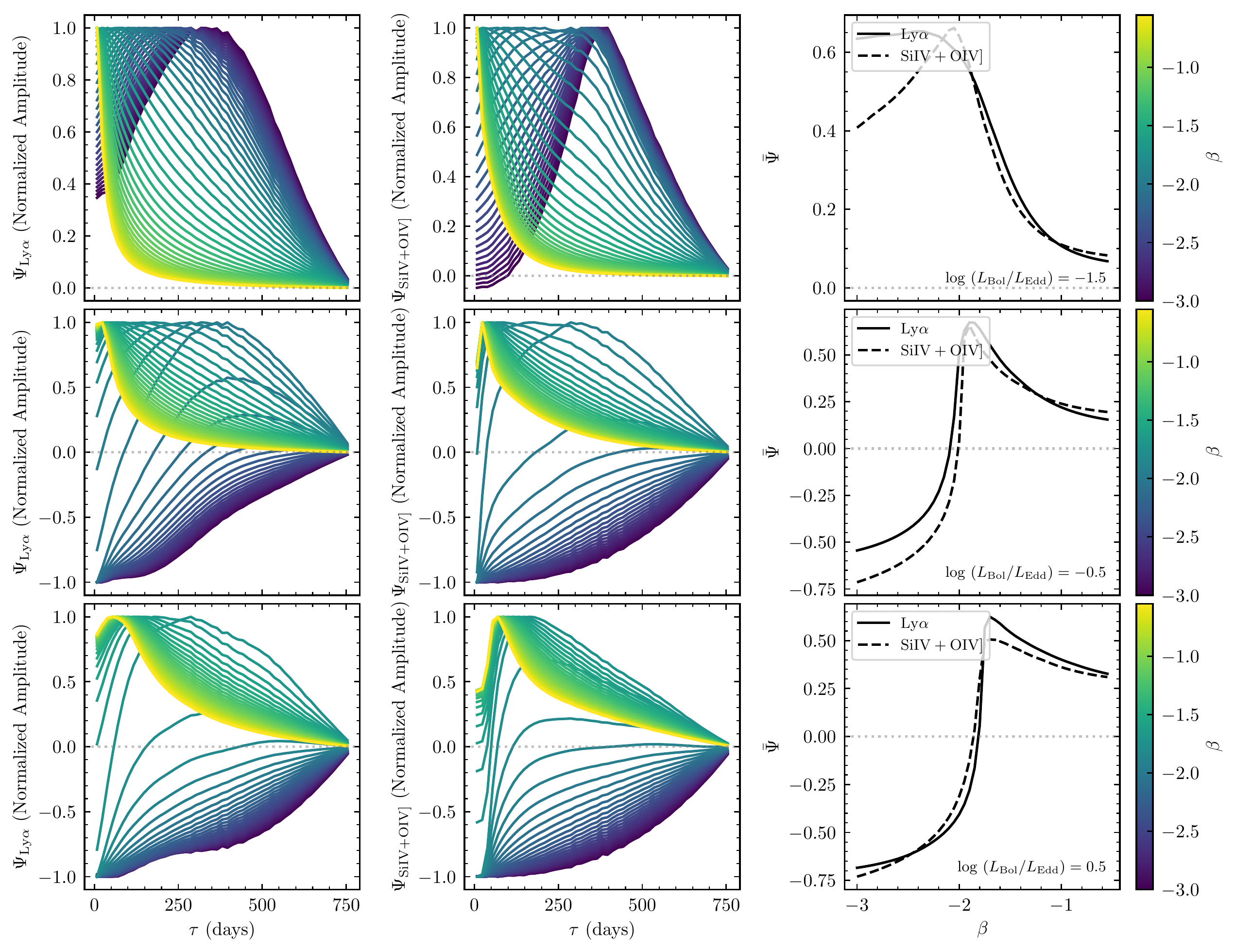}
    \caption{One-dimensional transfer functions $\Psi$ of the emission lines for
    different $\beta$ (Configuration B). The meanings of the panels, colors, and
    lines are the same as Figure \ref{fig:transfer_functions}.
    \label{fig:transfer_functions_highaccr}}
\end{figure*}

\begin{figure*}
    \figurenum{\ref{fig:transfer_functions_highaccr}}
    \centering
    \includegraphics[width=\textwidth]{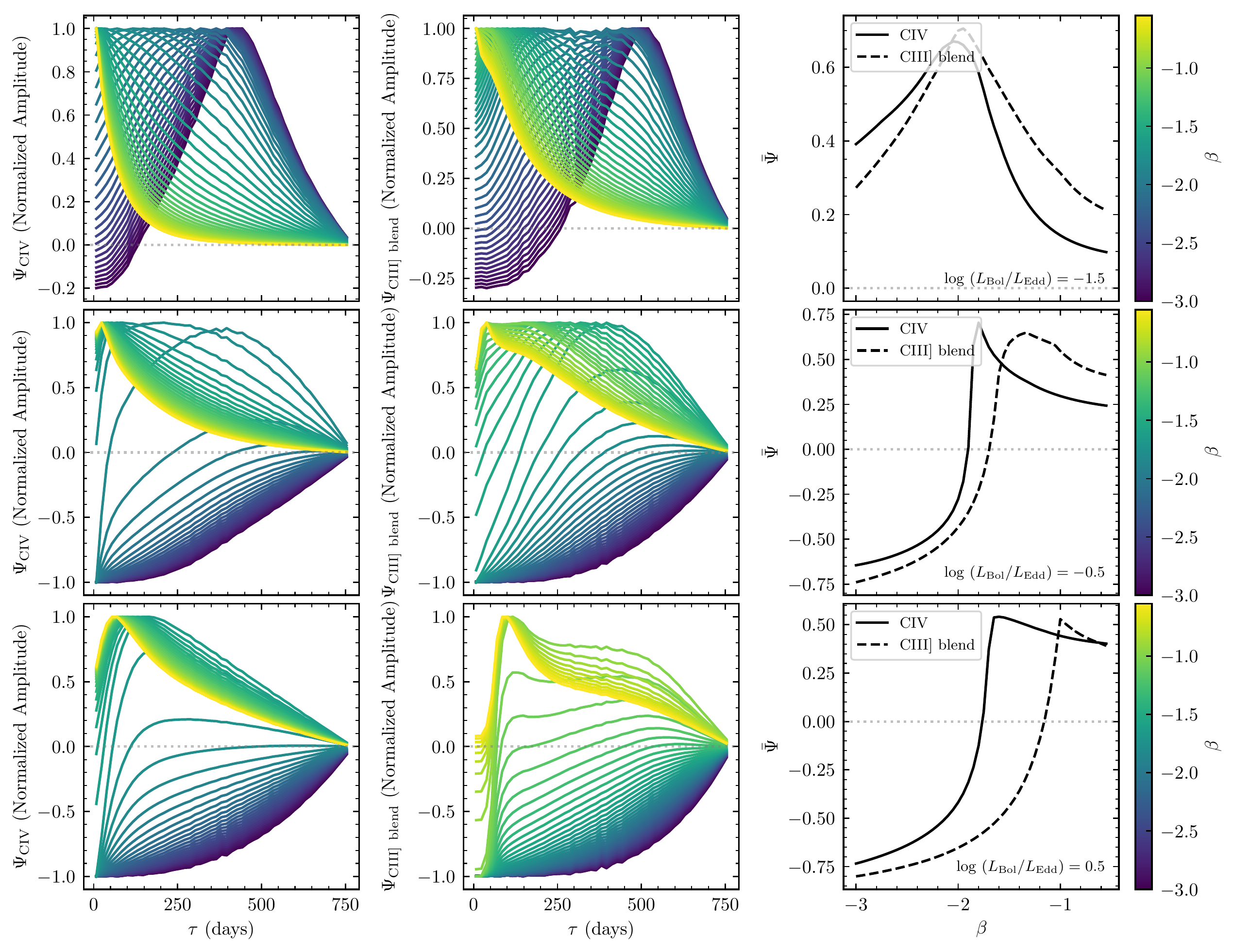}
    \caption{(Continued.)}
\end{figure*}

\begin{figure*}
    \figurenum{\ref{fig:transfer_functions_highaccr}}
    \centering
    \includegraphics[width=\textwidth]{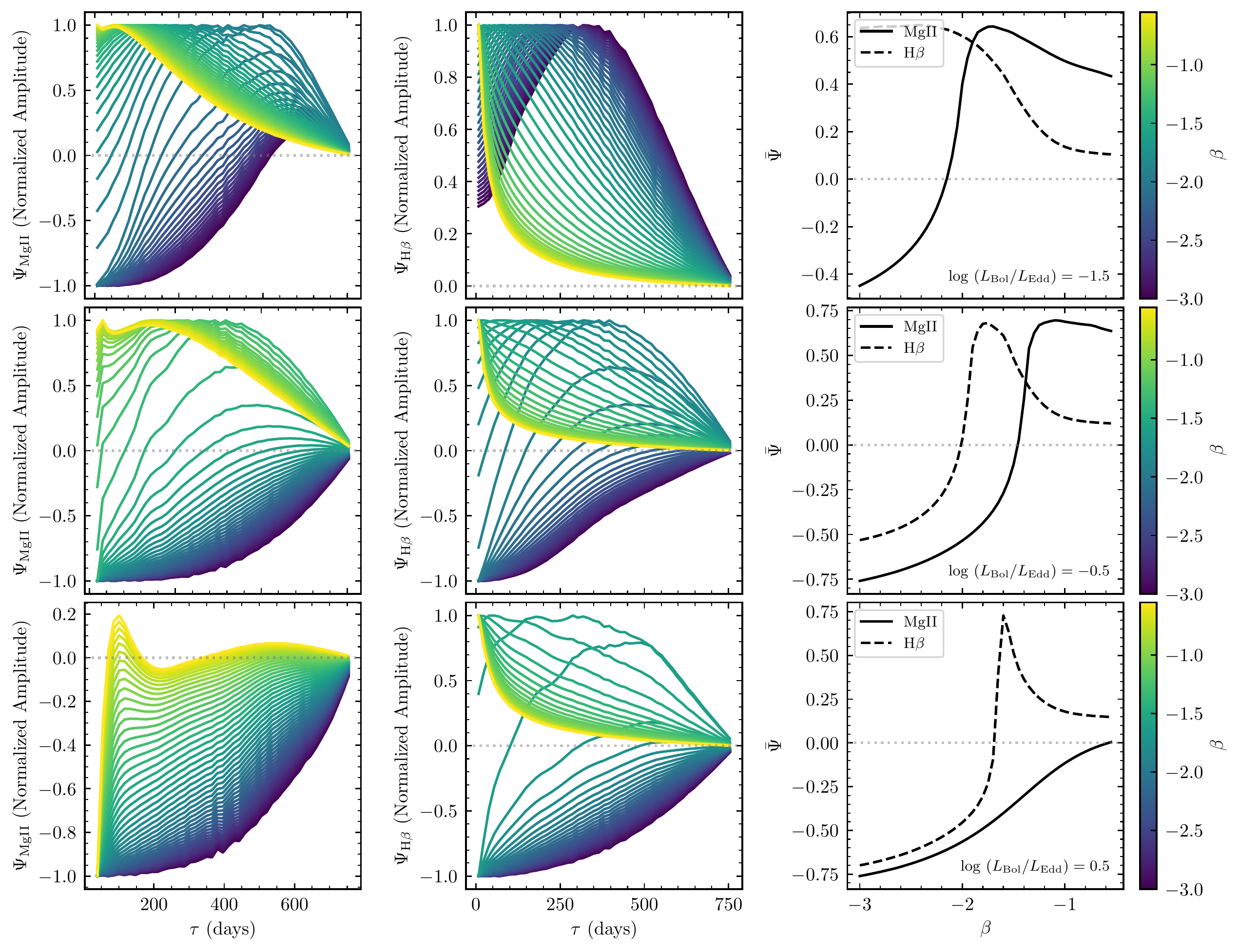}
    \caption{(Continued.)}
\end{figure*}

\begin{figure*}
    \centering
    \includegraphics[width=0.48\textwidth]{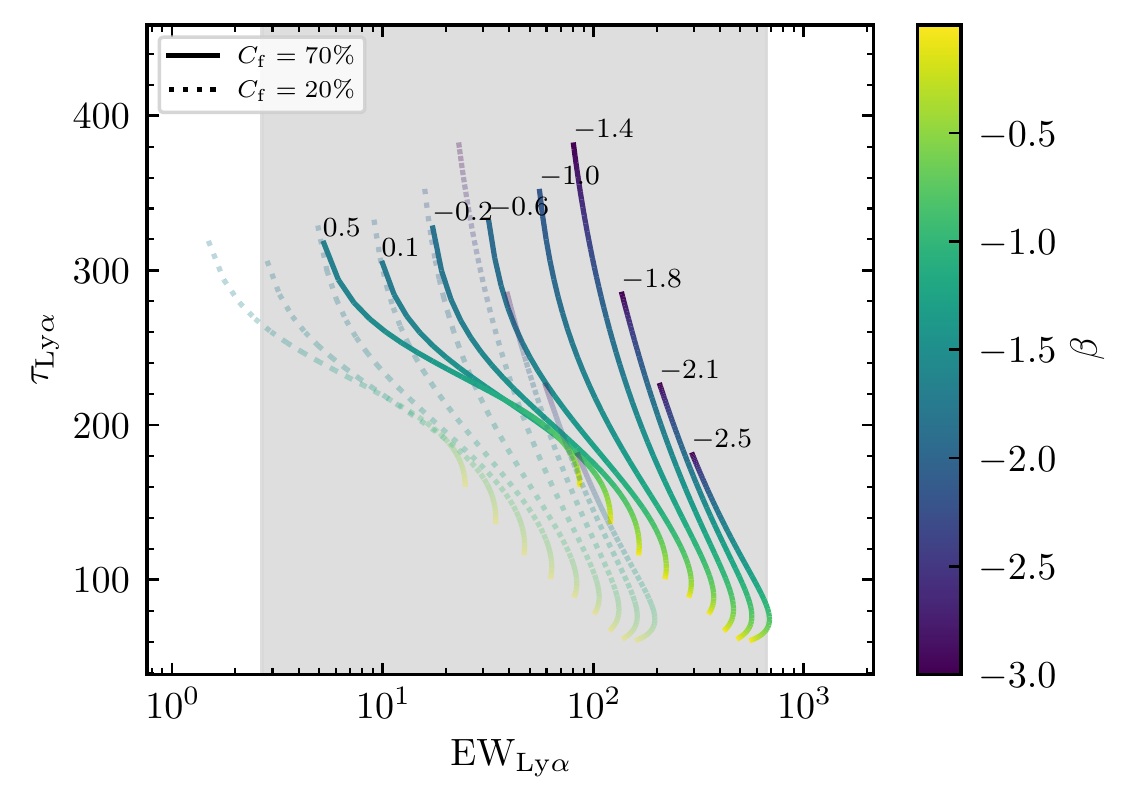}
    \includegraphics[width=0.48\textwidth]{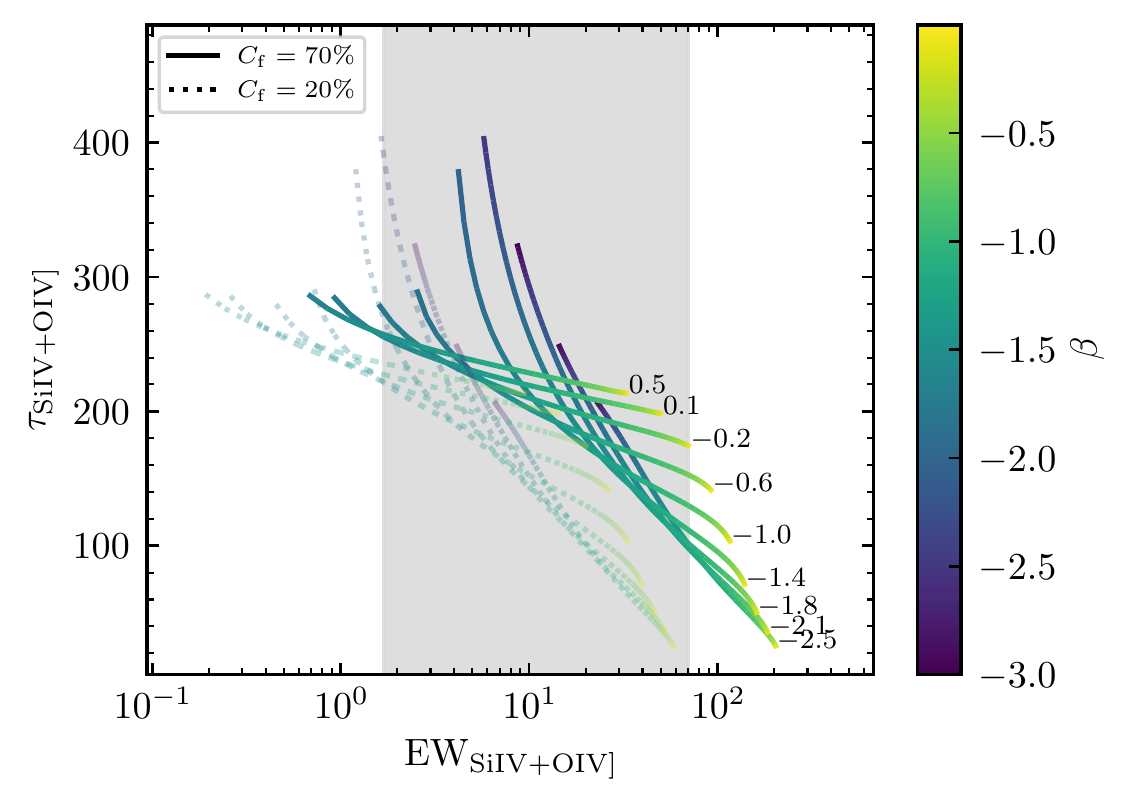} \\
    \includegraphics[width=0.48\textwidth]{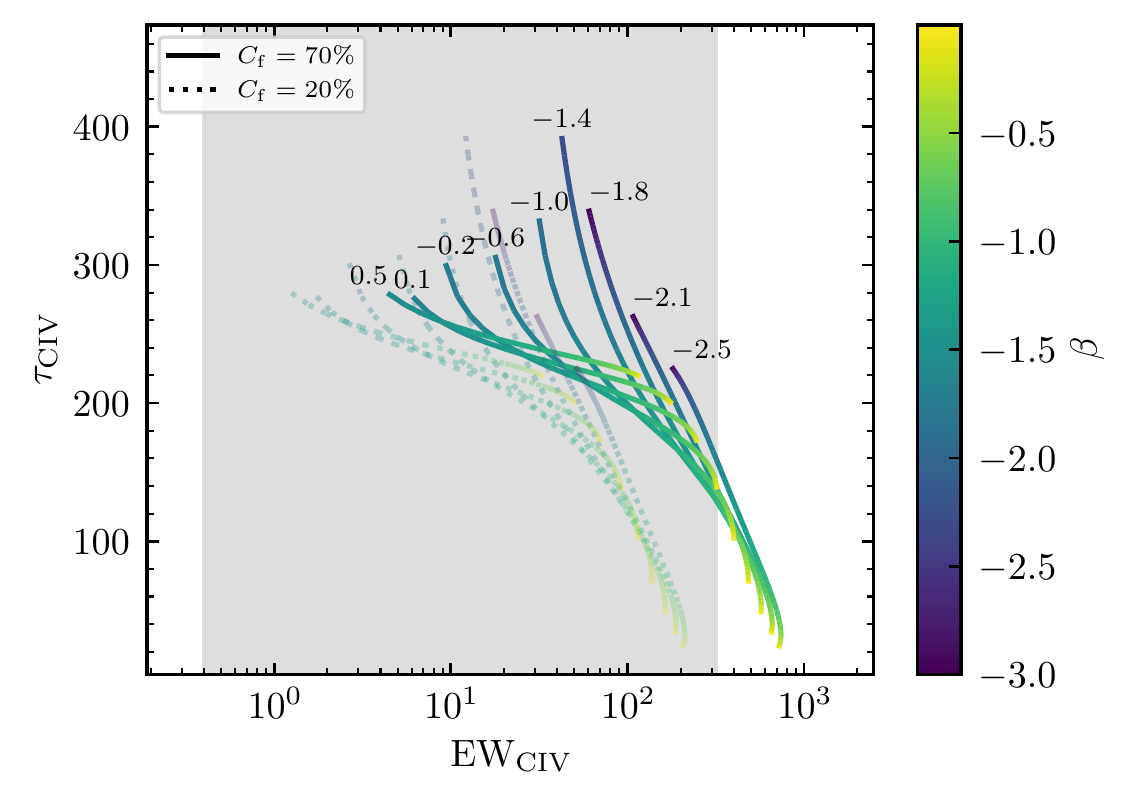}
    \includegraphics[width=0.48\textwidth]{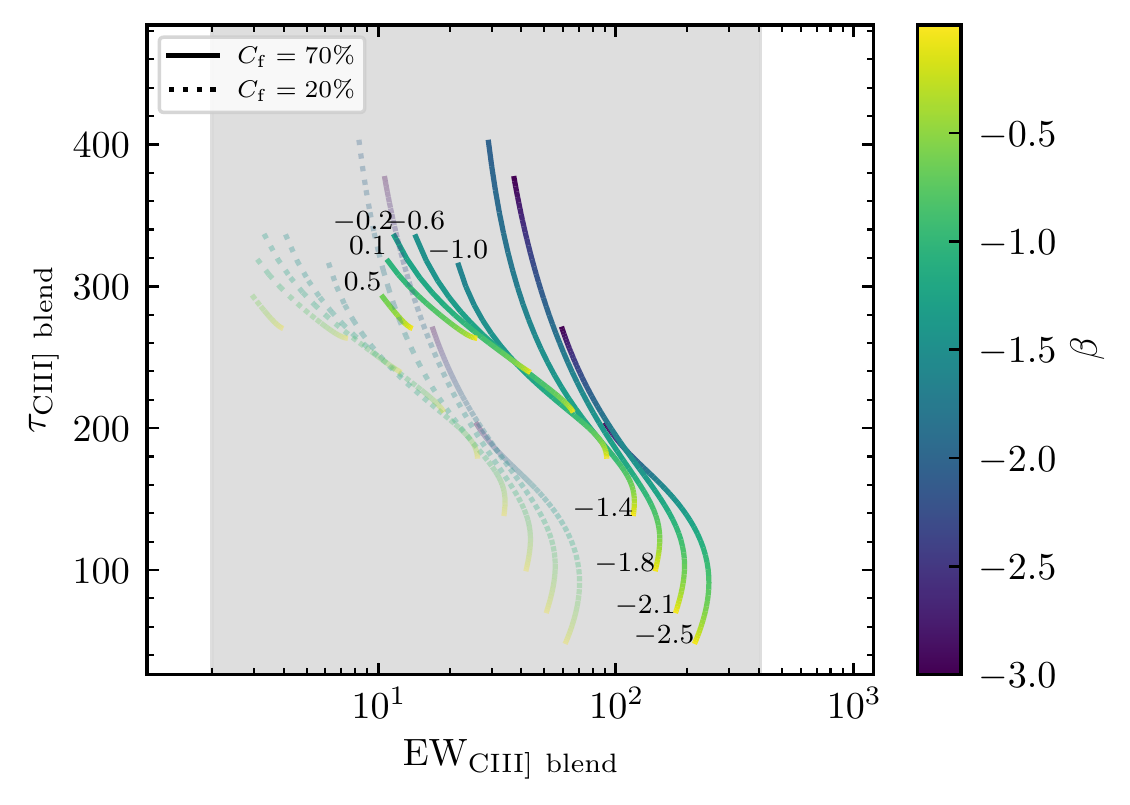} \\
    \includegraphics[width=0.48\textwidth]{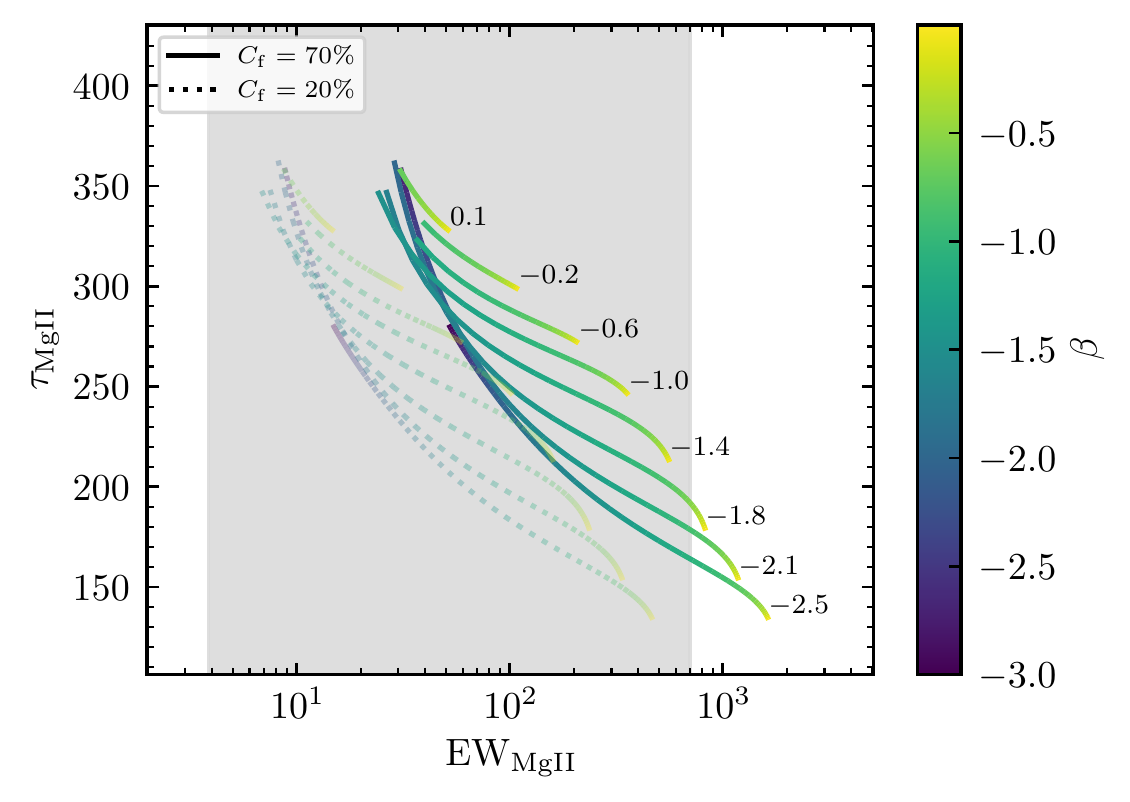}
    \includegraphics[width=0.48\textwidth]{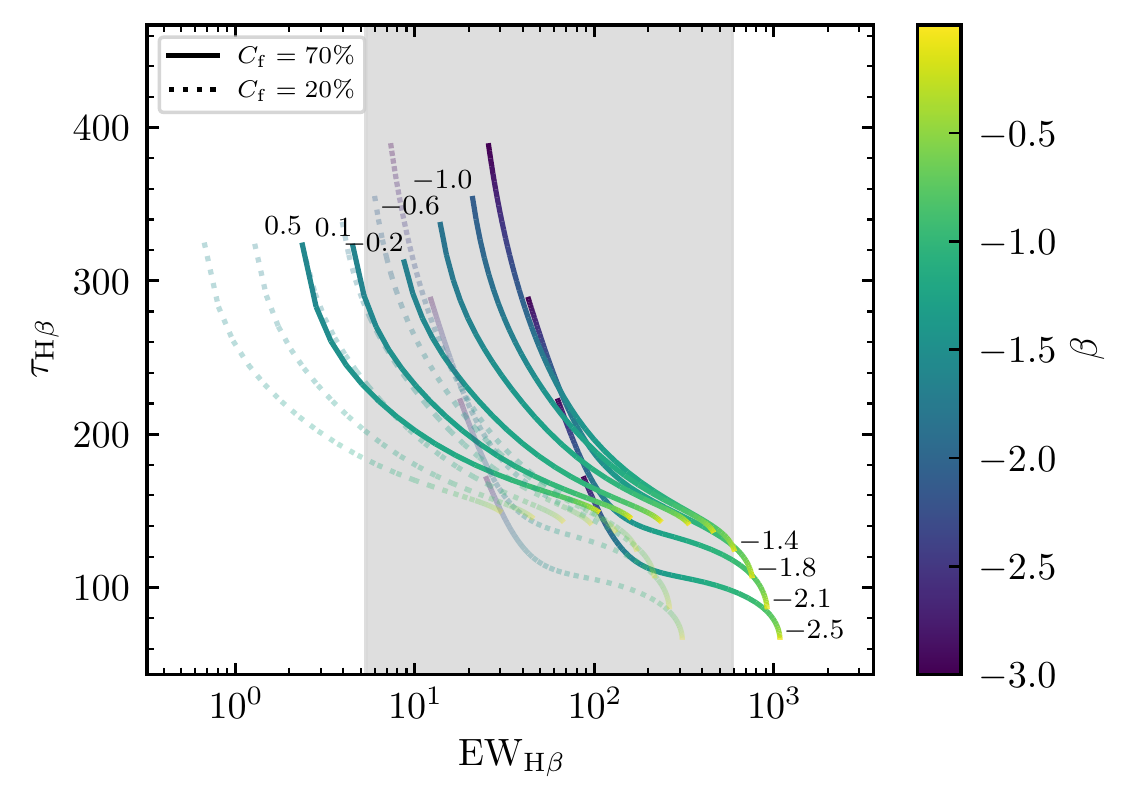}
    \caption{The correlations between EWs and time lags (Configuration B). The
    meanings of the panels, colors, and lines are the same as Figure
    \ref{fig:ew_lag}. \label{fig:ew_lag_highaccr}}
\end{figure*}

\begin{figure*}
    \centering
    \includegraphics[width=0.48\textwidth]{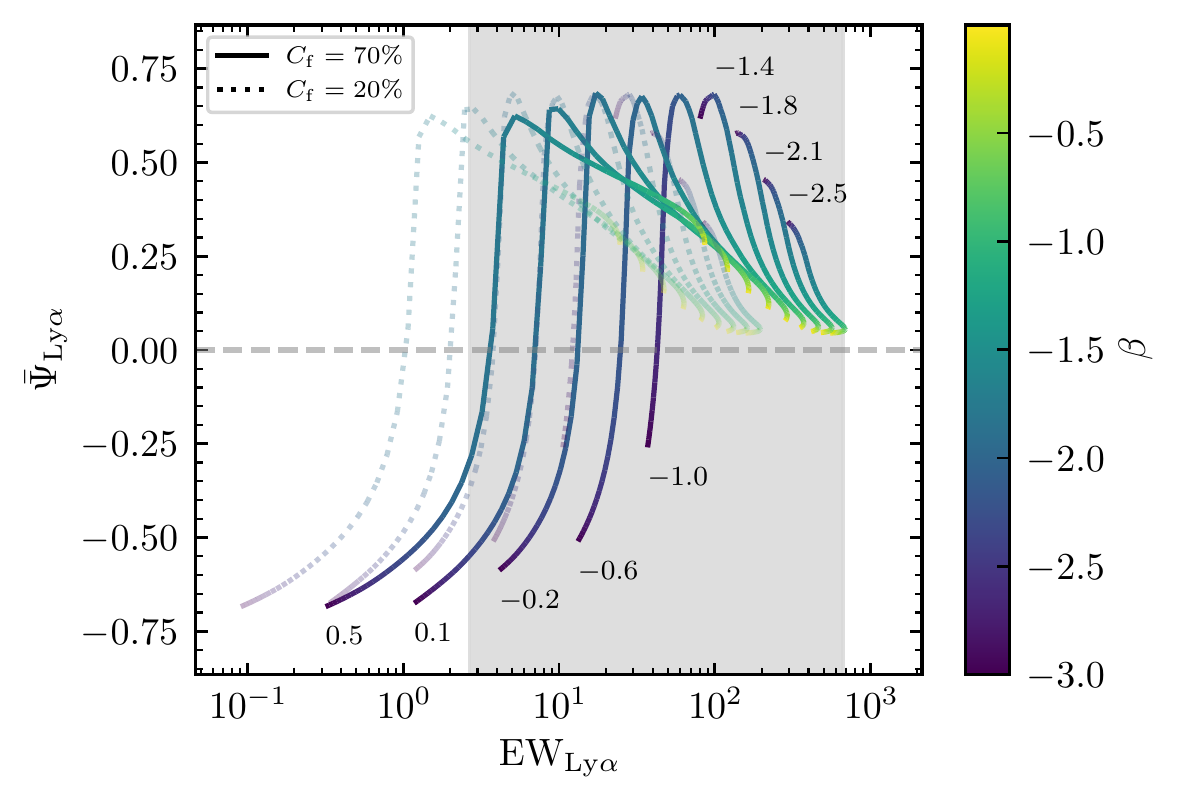}
    \includegraphics[width=0.48\textwidth]{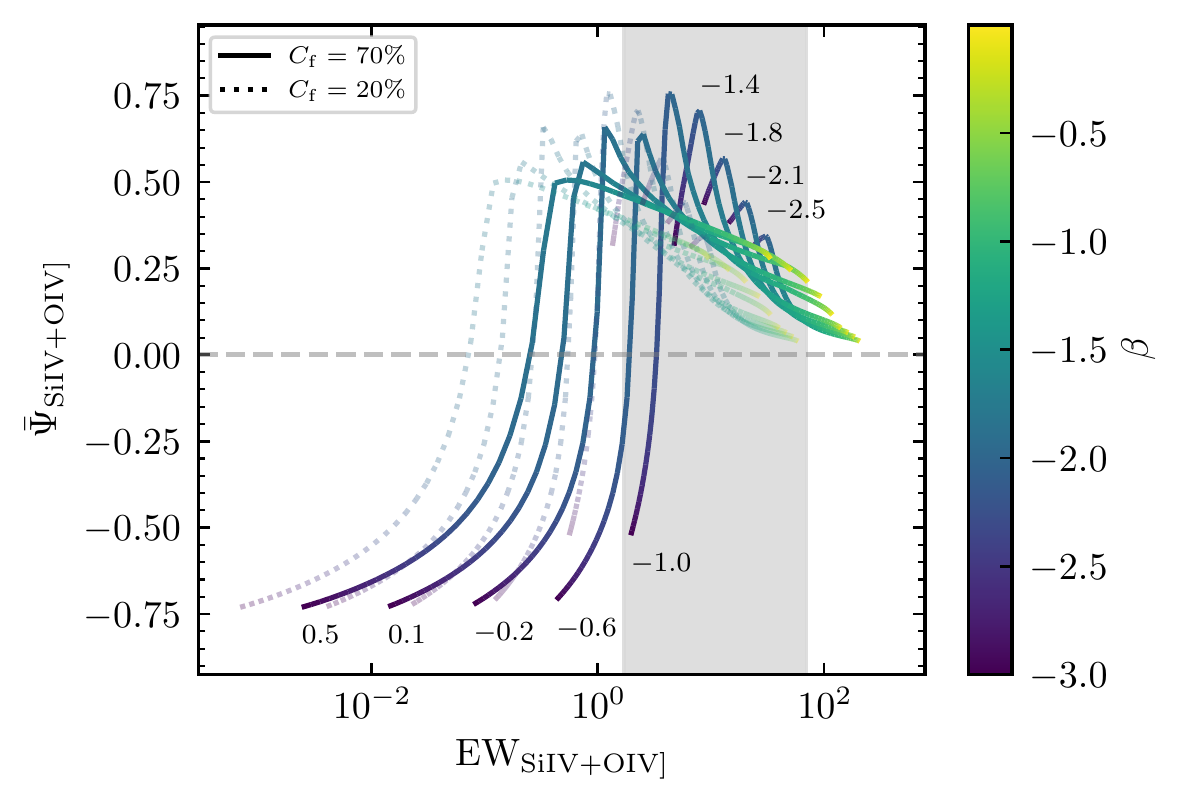} \\
    \includegraphics[width=0.48\textwidth]{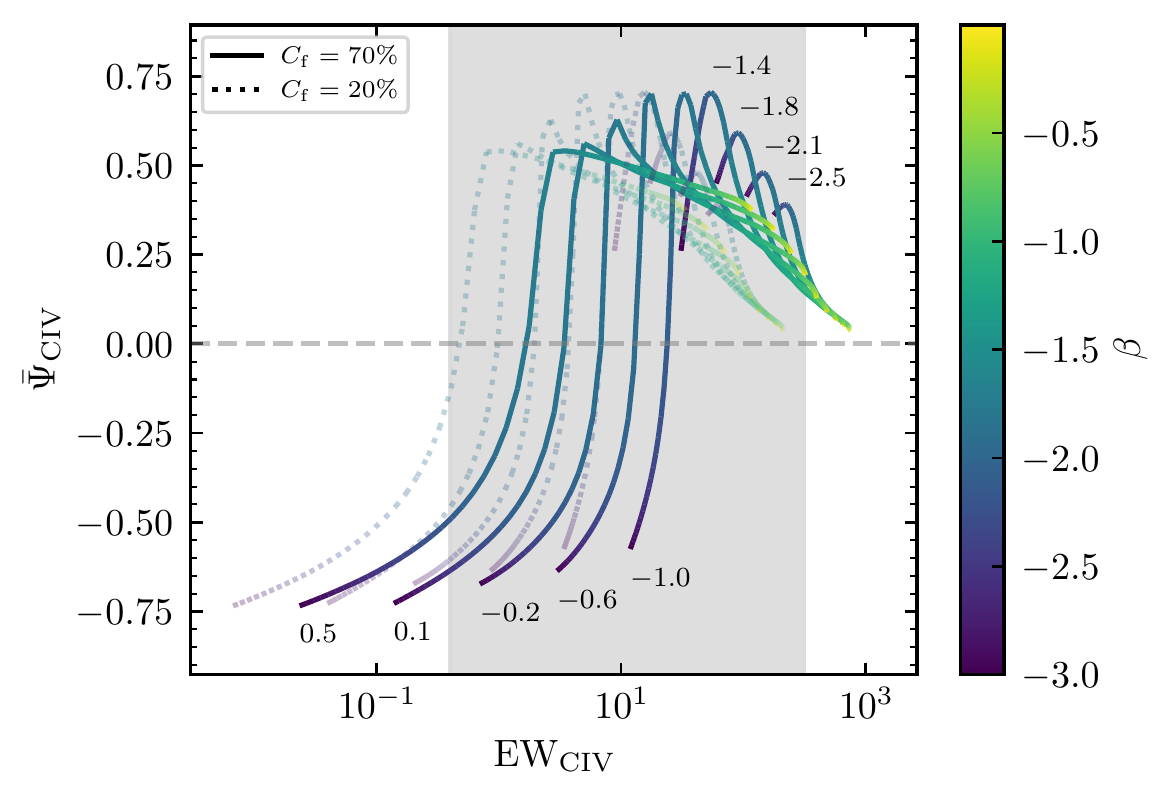}
    \includegraphics[width=0.48\textwidth]{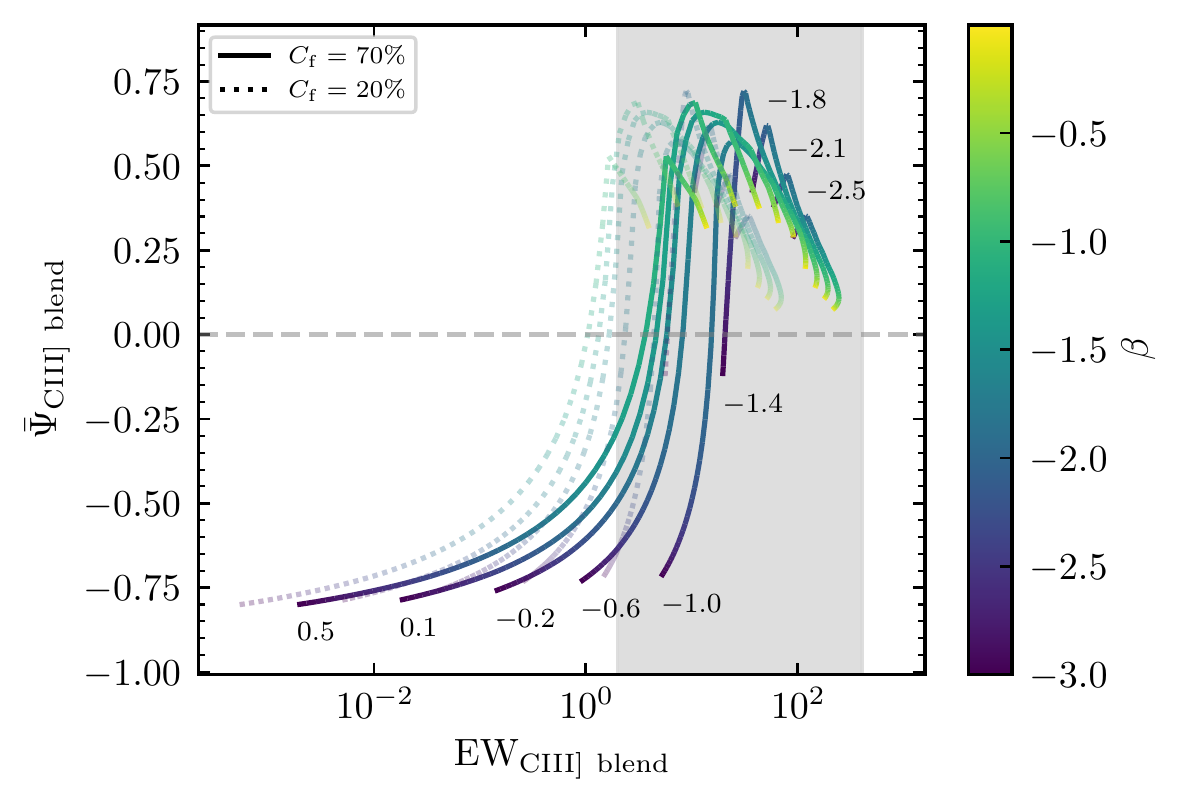} \\
    \includegraphics[width=0.48\textwidth]{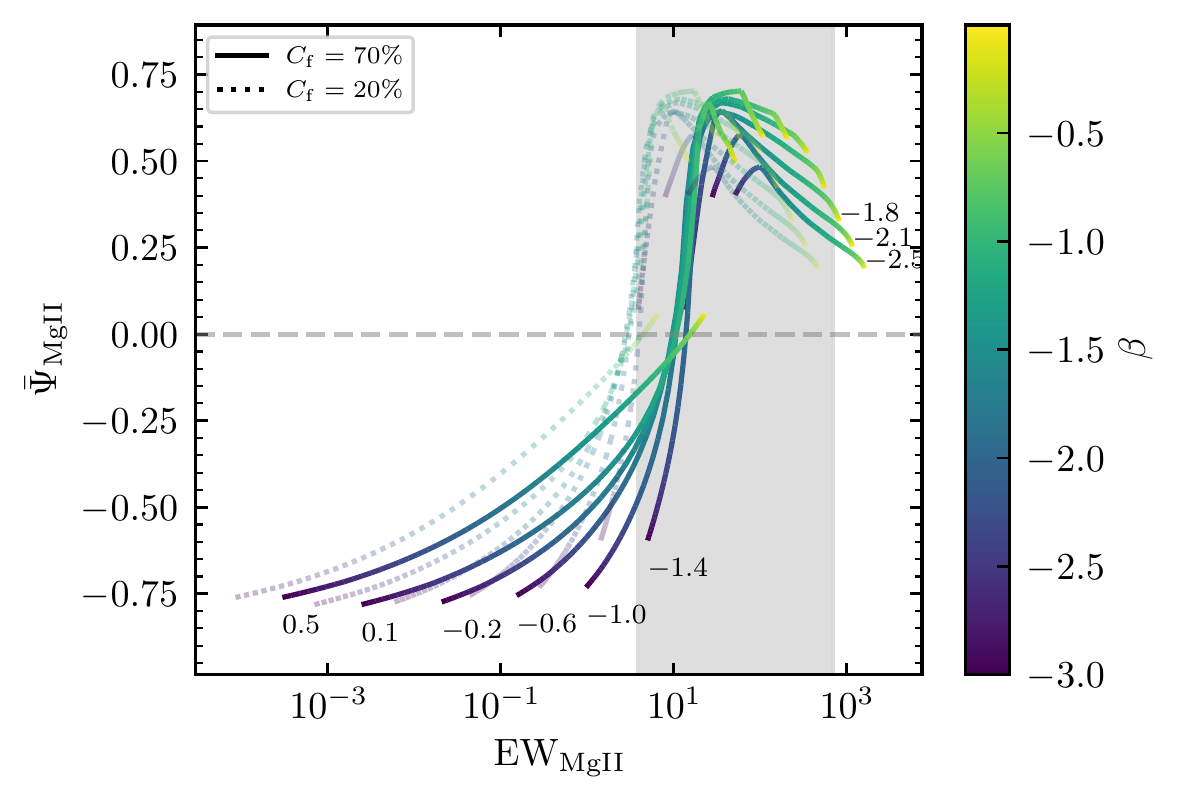}
    \includegraphics[width=0.48\textwidth]{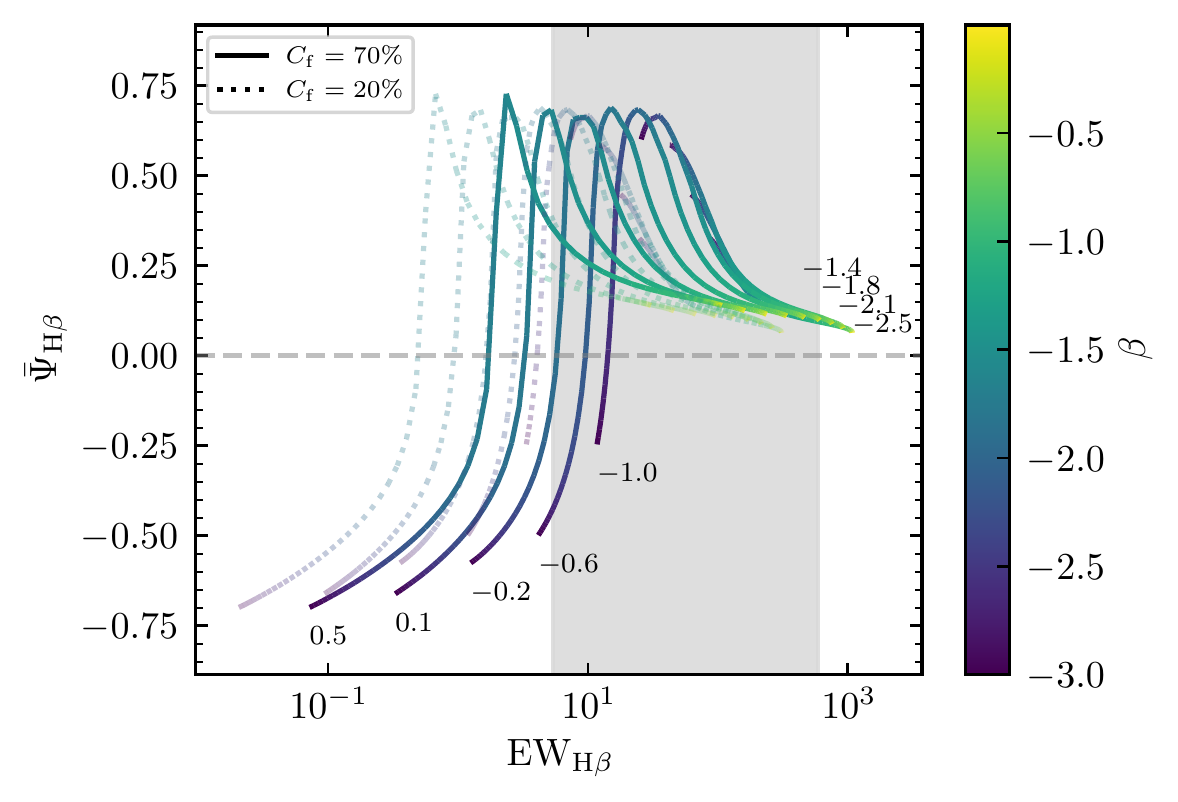}
    \caption{The correlations between EWs and the amplitudes of transfer
    functions (Configuration B). The meanings of the panels, colors, and lines
    are the same as Figure \ref{fig:ew_amplitude}.
    \label{fig:ew_amplitude_highaccr}}
\end{figure*}

\end{document}